\documentclass[preprint]{aastex}

\begin{document}

\def\K{{\rm\,K}}

\title{Preheated Advection Dominated Accretion Flow}

\author{Myeong-Gu Park}
\affil{Department of Astronomy and Atmospheric Sciences,
                 Kyungpook National University, Taegu 702-701, KOREA}
\email{mgp@knu.ac.kr}
\and
\author{Jeremiah P. Ostriker}
\affil{Princeton University Observatory, Princeton University,
       Princeton, NJ 08544}
\email{jpo@astro.princeton.edu}

\shorttitle{Preheated ADAF}
\shortauthors{Park \& Ostriker}

\begin{abstract}
The advection dominated accretion flow (ADAF) has been quite successful in
explaining a wide variety of accretion-powered astronomical sources.
The physical characteristics of ADAF complement the classical thin disk
flow quite nicely, and extensive work has been done on it. 
However, all high temperature accretion solutions including ADAF are physically
thick, so outgoing radiation interacts with the incoming flow. Thus, ADAF
solutions share as much or more resemblance with classical spherical accretion
flows as with disk flows. This interaction, which has been 
neglected by most authors, is primarily through the Compton
heating process which will typically limit the steady solutions to $L \la
10^{-2} L_{E}$, where $L_{E}$ is the Eddington luminosity. 
We examine this interaction for the popular ADAF (Narayan \& Yi 1995a) case but
similar conclusions, we expect, would apply for other high temperature,
geometrically thick disks as well. We study the global thermal nature
of the flow, with special consideration given to various cooling and,
especially, preheating by Comptonizing hot photons produced at smaller
radii. We find that without allowance for Compton preheating, a very restricted
domain of ADAF solution is permitted and with Compton preheating included a new
high temperature PADAF branch appears in the solution space.
In the absence of preheating, high temperature flows do not exist when the mass
accretion rate $\dot m \equiv \dot M c^2 / L_{E} \ga 10^{-1.5}$.
Below this mass accretion rate, a roughly conical
region around the hole cannot sustain high temperature ions and
electrons for all flows having $\dot m \ga 10^{-4}$, which
may lead to a funnel possibly filled with a tenuous hot outgoing wind.
If the flow starts at large radii with the usual equilibrium
temperature $\sim 10^4\K$, the critical mass accretion rate is much
lower, $\dot m \sim 10^{-3.7}$ above which level no self-consistent ADAF
(without preheating) can exist.
However, above this critical mass accretion rate, the
flow can be self-consistently maintained at high temperature if
Compton preheating is considered. These solutions constitute a new
branch of solutions as in spherical accretion flows. High temperature
PADAF flows can exist above the critical mass accretion rate in addition
to the usual cold thin disk solutions. Therefore, the Compton preheating could
be the mechanism to trigger the phase change from the thin disk to PADAF.
We also find solutions where the flow near the equatorial plane
accretes normally while the flow near the pole is overheated by
Compton preheating, possibly becoming, a polar wind, solutions which we
designate WADAF.

\end{abstract}


\section{Introduction}

Gas accretes onto compact objects in disk or (quasi) spherical shape
depending on the angular momentum and the entropy of the gas.
The flow may have a thin disk shape with very small radial motion,
or it could be spherical or spheroidal with significant radial motion.
The amount of radiation emitted depends on the shape of the flow
as well as on the nature of the compact object. Accretion onto white dwarfs and
neutron stars always produces luminosity approximately proportional to the mass
accretion rate. Thin disk accretion also produces radiation that is a constant
fraction of the gravitational energy. However, spheroidal or spherical flow
onto the black holes with relatively large radial velocities emits
differing amount of radiation, roughly scaling with the square of the accretion
rate (Shapiro 1973a; Park 1990a,b), depending on the physical states of the
gas and how the radiation escapes from the flow. Therefore, accretion
onto black holes can have a variety of geometrical flow shapes and can
produce a wide range of luminosity and emitted spectrum.

\subsection{Spherical accretion}

Accretion flows onto black holes become spherical or nearly spherical
when the angular momentum is very small or when the hole is rapidly
moving relative to the gas (Hoyle \& Lyttleton 1939; Bondi \& Hoyle 1944; Bondi
1952; Loeb \& Laor 1992). We review first the simpler case of spherical
accretion because we find that the type of solution obtained here are quite
relevant to more complex ADAF case. 
As long as two-body processes are important, the solutions become
scale-free and are applicable to black holes of arbitrary mass (Chang
\& Ostriker 1985). These scale-free solutions then depend primarily
on the dimensionless luminosity
\begin{equation}
l \equiv \frac{L}{L_{E}}
\end{equation}
and the dimensionless mass accretion rate\footnote{We define the Eddington mass
accretion rate as 
$ {\dot{M}_{E}} \equiv {L_{E}}/c^2 $, without $e^{-1}$ factor because the
radiation efficiency $e$ of accretion onto black holes is not fixed in
general.} 
\begin{equation}
\dot m \equiv \frac{\dot{M}}{\dot{M}_{E}}
\equiv \frac{\dot{M}c^2}{L_{E}},
\end{equation}
or, equivalently, the dimensionless luminosity $l$ and the radiation efficiency
\begin{equation}
e \equiv \frac{L}{\dot{M} c^2},
\end{equation}
where $L_{E}$ is the Eddington luminosity (Ostriker et al. 1976).
However, if we require the self-consistency between the gas and the
radiation field generated by the accretion process, only certain
combinations of the accretion rate and the luminosity (plus spectrum)
are allowed, and a given {\it self-consistent} solution is described by a line
in the 
$(l,\dot m)$ plane. Within a given type of solution, only one parameter
is needed to characterize the specific flow, and the dimensionless mass
accretion rate, $\dot m$, is the most meaningful parameter. Broad reviews of
accretion physics can be found, for example, in Rybicki \& Lightman (1979),
Pringle (1981), Treves, Maraschi, \& Abramowicz (1988), Frank, King, \& Raine
(1992), Chakrabarti (1996b).

\subsubsection{Spherical flow with small mass accretion rate ($\dot m \ll 1$)}

The electron scattering optical depth from infinity down to the horizon
is roughly given by $ \tau_{es} = {\dot m} (r / r_s)^{-1/2} $ for
freely falling flow where $r_s $ is the Schwarzschild radius of the hole and,
therefore, for $\dot m \ll 1$, spherical flow is optically thin to
scattering. Photons escape through the flow almost freely,
and the exchange of momentum and energy between the gas and the
radiation field is negligible. Radiative cooling and Compton heating
are also quite negligible. Gas is almost freely falling, $v/c \simeq
(r/r_s)^{-1/2}$, which is true even in relativistic flow (Shapiro
1973a,b, 1974); adiabatic heating keeps the gas up to the virial temperature,
$T \propto (r/r_s)^{-1}$, until electrons become relativistic (Shapiro
1973a; Park 1990a,b). Once electrons reach relativistic
temperature, the adiabatic index of the gas changes from $\gamma = 5/3$
to $\gamma = 13/9$ if electrons and protons are well coupled. Now the
temperature is increasing less rapidly and radiative cooling, mainly a
relativistic bremsstrahlung and synchrotron, becomes more efficient.
Regardless, the gas can reach as high as a few times $10^{10} \K$ even
when electrons are loosely coupled to ions via Coulomb process. The
luminosity and efficiency of the non magnetic solutions are quite small due to
the very low gas density, typically $l < 10^{-7}$ and $e < 10^{-6}$ ({\sl solid
line} marked as S [Shapiro 1973a] and {\sl large crosses} [Park 1990b]
in Figs. 1 \& 3) with efficiency approximately proportional to
the mass accretion rate $e \propto \dot{m}$ and luminosity $l \propto {\dot
m}^2$. The radiation temperatures $T_X$, defined as $4k T_X$ being equal to the
energy-weighted mean photon energy, of the solutions are plotted as
{\sl crosses} in Figure 2.

\subsubsection{Spherical flow with large mass accretion rate ($\dot m
\gg 1$)}

As the mass accretion rate $\dot m$ approaches 1, radiative cooling
becomes more efficient due to the increased density and the flow cools
down to $\sim 10^4\K$ in the absence of Compton heating. The flow forms
an effectively optically thick, i.e., optically thick to
absorption, core with blackbody radiation field inside. The flow no
longer behaves as an adiabatic gas.

Further increase of the mass accretion rate makes the core grow larger, and
soon the whole flow becomes effectively optically thick. The flow is
now cooled by the diffusion of radiation as in stellar interior
(Flammang 1982, 1984; Soffel 1982; Blondin 1986; Park 1990a; Nobili, Turolla,
\& Zampieri 1991). However, the ultimate source of the luminosity is gravity
through adiabatic compression rather than nuclear burning. {\sl
Triangles} in Figures 1--3 show the luminosity $l$, the radiation
temperature $T_X$, and the radiation
efficiency $e$ as a function of the mass accretion rate $\dot m$
(Nobili et al. 1991). Note that the high temperature, low $\dot m$
solutions ({\sl large crosses}) are quite distinct from the low temperature,
high $\dot m$ solutions ({\sl triangles}) in Figure 2, but
they are continuous in Figure 1. Although the luminosity increases
steadily with larger mass accretion rate, the flow is still a quite inefficient
radiator, $e \lesssim 10^{-7}$. 

The electron scattering optical depth becomes larger than unity when $\dot
m > 1 $, and the flow becomes optically thick to scattering. Photons
collide more often with the flow and the diffusion speed of photons
decreases with increasing optical depth. For high $\dot m$, the bulk velocity
of the flow exceeds the photon diffusion speed, $ v >
c/(\tau_{es}+\tau_{abs}) $, and radiation trapping occurs (Begelman
1978). The behavior of radiation and gas in this regime has to be
treated under rigorous relativistic framework (Flammang 1982, 1984; Park
1990a; Nobili et al. 1991; Park 1993). In such an optically thick
relativistic flow, the flux seen by the stationary observer relative to
background spacetime is significantly different from the flux seen by
the observer comoving with the flow. The momentum transferred to the
gas from the radiation is closely linked to the former, a
comoving-frame flux, and the luminosity seen by the observer at
infinity to the latter, a fixed-frame flux. (Mihalas and Mihalas 1984;
Park 1993). Also in this regime, the usual diffusion relation between
the flux and the radiation pressure does not hold and should not be
used (Miller 1990; Park \& Miller 1991). This radiation trapping
critically affects all types of accretion flow with large mass
accretion rates.

The possibility of another mode at the same mass accretion rate was
first investigated by Wandel, Yahil, \& Milgrom (1984). Subsequent
rigorous studies show that the flow can be maintained up to high
temperature through Compton preheating by the radiation that accretion
flow itself produces at smaller radii (Park 1990a,b; Nobili et al.
1991; Mason \& Turolla 1992). They find this high-temperature,
higher-luminosity, and 
therefore higher-efficiency flow does exist, but only for $ 3
~\lesssim~ \dot m ~\lesssim~ 100$ as shown in Figures 1--3 as {\sl
large open circles} (Park 1990b). Hence, for a
certain range of the mass accretion rate, there exists more than one
family of accretion flow with different properties.

These high-temperature accretion flows have the gas temperature as
high as $10^9 \sim 10^{10} \K$ in contrast to the low-temperature ($10^4 \K$)
case. The 
flow is optically thick to scattering, and radiation is trapped inside
$r \simeq \dot m r_s$. Yet, the flow is optically thin to absorption
despite the large scattering optical depth because of the high electron
temperature. The flows are much more luminous $l ~\simeq~
10^{-4}-10^{-2}$ ({\sl large open circles} in Fig. 1) than their
low-temperature counterparts ({\sl triangles}) and produce much harder
photons. The
radiation efficiency $e \sim 10^{-4}$ is about $10^3$ times higher than
for the low-temperature flow, yet still far less than $e \simeq 0.1$ of the
thin disk. The self-consistent luminosity $l$ increases with the mass
accretion rate $\dot m$, while the spectrum of the emitted radiation
which preheats the flow in turn is harder for lower $\dot m$ and softer
for higher $\dot m$ ({\sl circles} in Fig. 2). The global stabilities of these
flows have 
been investigated in relativistic framework, and they are found to be
thermally unstable and develop moving shocks (Zampieri, Miller, \&
Turolla. 1996) as suspected in steady-state work (Park 1990a,b).

There exist a few candidate processes that might significantly increase the
radiation efficiency. M\'esz\'aros (1975) considered the dissipational heating
from turbulent motion, Maraschi et al. (1982) magnetic field
reconnection, and Park \& Ostriker (1989) electron-positron pair
production. Solutions of the first and second types can reach $e$ as
high as $\sim 0.1$, {\sl small squares} in Figures 1 \& 3 (Maraschi et
al. 1982), while that of the last type up to $e \sim 10^{-2.5}$, {\sl
small filled circles} in Figures 1 \& 3 (Park \& Ostriker 1989). They
constitute yet another type of accretion flow in this so called
super-Eddington accretion regime. However, the foundations for these
higher-efficiency families are less solid than the other two families
because the solutions are found without the consideration of preheating
(M\'esz\'aros 1975; Maraschi et al. 1982) or only the inner supersonic
part of the flow is dynamically considered (Park \& Ostriker 1989). The
rigorous and full treatment of dynamics and preheating will certainly
show that they are unstable to the preheating instability (Cowie, Ostriker, \&
Stark 1978; Ciotti \& Ostriker 1999). It is possible
that the overheated flow is stabilized by developing a shock (Chang \&
Ostriker 1989; {\sl small crosses} in Figs. 1 \& 3), yet no
self-consistent solutions with preheating and a steady shock have been
constructed despite some hints of their existence (Park 1990b; Nobili et
al. 1991).

\subsection{Preheating}

For this higher $\dot m$ and $l$ flow, radiation has more chance to
interact with the gas: both momentum and energy are transferred from
one to the other. Transfer of momentum, when the interaction of radiation with
matter is mediated by the electron Thomson scattering, leads to the critical
Eddington luminosity ($l=1$ lines in Figs. 1 \& 3). 
Another lower type of luminosity upper limit is found by Ostriker et al.
(1976) that can be as small as $10^{-2}$ times the Eddington
luminosity. This new critical luminosity exists due to preheating of
gas flow by hard Compton radiation. When the gas is preheated
significantly near the sonic point, it becomes too hot to accrete in a
steady state. For example, when the surrounding gas temperature is near
$10^4 \K$ and the Compton temperature of preheating radiation is $10^8
\K$, steady-state accretion is impossible above $l_{cr} \simeq 0.01$
({\sl rectangular region I} in Figs. 1 \& 3). This preheating
explains why self-consistent solutions are not found for $\dot m
\lesssim 3$ and $\dot m \gtrsim 100$ (Park 1990a,b; Nobili et al.
1991): since Compton preheating depends on the energy and the
luminosity of outcoming photons, the preheating instability occurs when the
radiation is too hard ($\dot m \lesssim 3$; Fig. 2) or the luminosity is too
high ($\dot m \gtrsim 100$; Fig. 1). The region can be smaller
under certain 
conditions (Ostriker et al. 1976; Cowie et al. 1978;
Bisnovatyi-Kogan \& Blinikov 1980; Stellingwerf \& Buff 1982; Krolik \& London
1983). 

Preheated flow shows interesting time-dependent behavior (Cowie et al.
1978; Grindlay 1978) that may be applicable to variabilities of accretion
powered 
astronomical sources, including galactic X-ray sources, active galactic
nuclei, QSOs, and cooling flows (Ciotti \& Ostriker 1997, 1999).
Cowie et al. (1978) found two distinct type of
time-dependent behavior inside and near the sonic radius. The flow is
overheated inside the sonic point for high $l$, low $e$ accretion
({\sl region II} in Figs. 1 \& 3), producing recurrent flaring on short time
intervals. Low $l$, high $e$ flow ({\sl region III} in Figs. 1 \& 3) develops
preheating outside the sonic radius, causing accretion rate and
luminosity changes on longer time scales. However, these time-dependent
studies did not impose self-consistency, meaning the radiation
efficiency is arbitrarily assigned rather than calculated from the flow
itself. 

All of the high temperature, physically thick self-consistent solutions shown
in Figure 3 that are known or expected to be stable have a quite low efficiency
($e \la 10^{-4}$). It seems likely that all physically thick solutions
having a high enough density to efficiently produce radiation will undergo
relaxation oscillations with the primary cause being preheating
instabilities. The exception to this may be the slim disks which have a
relatively low emission temperature and which essentially resemble massive
stars.

\subsection{Axisymmetric accretion}

Accretion flow onto a black hole {\sl with} angular momentum can
assume disk-like or spheroidal shape depending on how efficiently the
heat produced is removed. The flow becomes thin disk-like when the energy
from viscous dissipation is readily radiated away in the vertical
direction (Shakura \& Sunyaev 1973). If too much radiation is produced
and trapped within the flow, the disk puffs up to a thick one, and much
of the radiation and gas energy is advected into the hole with
significant radial velocity (Jaroszy\'nski, Abramowicz, \& Paczy\'nski
1980; Paczy\'nski \& Wiita 1980). When the disk is not thin,
constructing the flow solution becomes a two-dimensional problem which
has been tackled only by arbitrarily fixing certain physical
parameters (e.g., Paczy\'nski 1998), or by reducing it to one-dimensional
problem (Abramowicz et al. 1988; Narayan \& Yi 1994, NY1 hereafter;
Narayan \& Yi 1995b, NY3 hereafter), or by numerical
simulations (Molteni, Lanzafame, \& Chakrabarti 1994; Ryu et al. 1995;
Chen et al. 1997; Igumenshchev \& Beloborodov 1997; Igumenshchev \& Abramowicz
1999; Stone, Pringle, \& Begelman 1999), or by assuming
self-similar forms (Narayan \& Yi 1995a, NY2 hereafter).

\subsubsection{Thin disk}

Since the seminal works of 70's, thin disk accretion, especially the
so-called $\alpha$-disk has been {\it de facto} standard accretion mode
applied to various accretion powered sources (Shakura 1972; Pringle \& Rees
1972; 
Shakura \& Sunyaev 1973; Novikov \& Thorne 1973). The thin disk has merits
of being simple: the equations are reduced to one-dimensional ones and
all physical interactions can be described by local quantities. Angular
momentum of the rotating gas is transported outward by the viscous
stress through the differentially rotating gas. Gravitational potential
energy is radiated away via viscous dissipation. Although the disk is
geometrically thin, it is optically thick to absorption, and radiation
diffuses out in the vertical direction. Because of the geometrical
shape, the emitted radiation does not in general interact with other parts of
the flow. The innermost edge of the disk determines the amount of
gravitational potential energy to be released, and the radiation
efficiency is roughly $0.1$ regardless of the central compact object being
neutron star or black hole. 
Therefore, all thin disk accretions appear
on {\sl the dotted line} $e=0.1$ (TD in Figs. 1 \& 3).

However, the thin disk formalism may not be extended to high luminosity
systems or, equally, high $\dot m$ systems: the disk becomes unstable
to thermal or secular instability when radiation pressure dominates
over the gas pressure (Lightman \& Eardley 1974; Shakura \& Sunyaev
1976; Pringle 1976; Piran 1978). Furthermore, hard X-ray sources like
Cygnus X-1 cannot be explained by a thin disk: it is too cool and
emits blackbody-like spectrum. Though Shapiro, Lightman, \& Eardley (1976)
found a new 
family of optically thin disk with very high electron temperature, they
are also proved to be unstable to the thermal instability (Pringle
1976; Piran 1978; Park 1995). Thus, for high flow rates or high temperature
flows, this simple, high efficiency solution is not available or appropriate.

\subsubsection{Thick disk}

As the dimensionless mass accretion rate $\dot m$ approaches $e^{-1}$ (i.e., $l
\rightarrow 1$),
the vertical height of the disk becomes comparable to the radius. In
the thin disk case, gravity is balanced by the Keplerian rotation
and the radial motion is negligible. When the disk becomes
geometrically thick, the dynamical balance in the radial direction becomes
important, and significant radial motion occurs, resulting in sub-Keplerian
rotation. Now the disk is two-dimensional: dynamics in the radial direction
as well as in the azimuthal direction should be considered.

This geometrically thick accretion disk has been studied by Paczy\'nski
and collaborators (e.g. Paczy\'nski \& Wiita 1980; Jaroszynski et al. 1980;
Abramowicz, Calvani, \& Nobili 1980). 
Thick disk flow can only be described by full two-dimensional partial
differential equations because the angular momentum of the gas is not known a
priori, radial infall motion can not be ignored, and the energy equations
become 
non-local. Studies of thick disk onto black holes show that the flow develops a
cuspy inner edge through which the gas flows into the hole with roughly
free-fall velocity. Photons are locked inside the disk and advected into the
hole; yet a large amount of radiation may escape through the surface of the
thick 
disk which shapes up the funnel around the rotation axis.

\subsubsection{Slim disk}

To eschew the complexities while retaining the essential features
of the thick accretion flow,
Abramowicz et al. (1988) applied height-integrated $\alpha$ disk
equations to super-Eddington ($\dot m > e^{-1}$) accretion flow. They
found stable disk solutions with significant radial motion in high
$\dot m$ regime, in which the thin disk is unstable. The flow is
optically thick and the radiation is trapped at small radii so that, as noted
above, these solutions resemble massive stars. The
entropy is directly advected with the flow, which adds to the radiative
cooling through the surface of the disk. This new family of disk
solutions is called 'slim disk' because the disk is assumed to be slim
enough to validate the height-integrated equations as in the thin disk
solutions. In this accretion mode the radiation efficiency is not
predetermined: some of the gravitational potential energy is ultimately
absorbed into the hole due to the entropy advection, and the amount of
radiation emitted depends on specific conditions of the flow. In this
regard, the slim disk has some similarity to the case of high $\dot m$
spherical accretion. The 
luminosity and radiation efficiency of these slim disks are plotted in
Figures 1 \& 3 as {\sl dot-dashed curve} labeled SD; the flow can reach
super-Eddington luminosity, $l \gtrsim 1$, and $e$ decreases from $0.1$
as $\dot m$ increases (Szuszkiewicz, Malkan, \& Abramowicz 1996). However, it
is not at all clear if the slim disk approach can be extended to
super-Eddington systems because the disk is likely to be thick rather than
slim. 

\subsubsection{Advection dominated accretion flow}

The slim disk solutions were constructed by explicitly integrating
the height-averaged hydrodynamic equations (Abramowicz et al. 1988).
However, Narayan \& Yi (NY1) find that these one-dimensional equations admit
simple self-similar solutions if cooling due to the entropy advection
is assumed to be a constant fraction of the viscous heating (also see
Spruit et al. 1987): The density, velocity, angular velocity, and total
pressure are simple power laws in radius: $\rho \propto r^{-3/2}$, $v
\propto r^{-1/2}$, $\Omega \propto r^{3/2}$, and $P \propto r^{-5/2}$.
Most of the dissipation energy generated is advected inward with the
flow, and the solutions are called advection dominated accretion flow
\footnote{The basic formulation of ADAF is the same as that of the slim disk
flow. They are different only in $\dot m$, or in the effective optical
depth.}
(ADAF; see Narayan, Mahadeva, \& Quataert 1998b for review). 
This type of accretion had been studied earlier (Ichimaru 1977) under different
names like `ion torus' (Rees et al. 1982).

However, since the temperature of this type of flow is close to the virial
value, it is not certain whether height-integrated equations are
adequate. In the subsequent work (NY2), the full
two-dimensional self-similar solutions have been constructed under the
standard $\alpha$-viscosity assumption. The radial velocity of the flow
turns out to be a fraction of (roughly proportional to $\alpha$ for
$\alpha \ll 1$) the free-fall value at the equatorial plane, 
yet the flow along the pole is pressure supported with zero infall
velocity. The matter is preferentially accreted along the equatorial
plane. The flow is subsonic due to the near virial pressure, which is valid for
$r \gg r_s$.  
Another unique feature of ADAF is that
the Bernoulli constant of the flow is positive, as noted by NY2 themselves and
critically analyzed by Blandford \& Begelman (1999), which has led Xu \& Chen
(1997) and Blandford \&
Begelman (1999) to propose a generalization of ADAF that has both inflow and
outflow at the same time. Another important result from NY2 is
that the physical quantities calculated from the vertically integrated
one-dimensional equations generally agree with those averaged over the
polar angle. In a way this justifies the slim disk approach to the
two-dimensional accretion flow, although this is proved only in the
self-similar flow. This permitted further extensive works on the
dynamics of ADAF to be based on the one-dimensional height-integrated
hydrodynamic equations (Narayan, Kato, \& Honma 1997; Chen, Abramowicz,
\& Lasota 1997; Nakamura et al. 1997; Gammie \& Popham 1998; Popham \&
Gammie 1998; Chakrabarti 1996a; Lu, Gu, \& Yuan 1999). 
Because, everywhere in the flow the radial velocities are quite substantial and
the pole to equator density variations are moderate, these solutions share many
of the characteristics of the purely spherical flow.

Detailed microphysics are readily incorporated into these one-dimensional
solutions, 
showing the main characteristics of ADAF: low radiation efficiency and
high electron temperature (NY3; Abramowicz et al.
1995). These make ADAF applicable to low luminosity, hard X-ray
sources, like Sgr A$^\ast$ (Narayan, Yi, \& Mahadevan 1995; Mahadevan,
Narayan, \& Krolik 1997; Manmoto, Mineshige, \& Kusunose 1997; Narayan
et al. 1998a), NGC 4258 (Lasota et al. 1996), soft X-ray transient
sources (Narayan, McClintock, \& Yi 1996; Narayan, Barret, \&
McClintock 1997), and low-luminosity galactic nuclei (Di Matteo \&
Fabian 1997; Mahadevan 1997) as well as to other diverse problems like
lithium production in AGN (Yi \& Narayan 1997), torque-reversal in
accretion-powered X-ray pulsars (Yi \& Wheeler 1998), the QSO evolution
(Yi 1996; Choi, Yang, \& Yi 1999), and X-ray background (Yi \& Boughn
1998). Constructing the 
accretion flow with the inner part being ADAF and the outer part being thin
disk has more merits and freedom than the simple thin disk, and therefore it
can be successful in representing a wider range of astronomical systems.

Typical ADAFs are presented as {\sl dashed lines} in Figures 1 \& 3 (NY3).
They are inefficient radiators: most of the gravitational energy of the
accreted matter is carried with the flow into the hole without
being released in the form of radiation. This very property of ADAF is
similar to that of the hot, adiabatic, low-$\dot m$ spherical accretion
flow (Shapiro 1973a,b) as the slim disk is to the low-temperature,
high-$\dot m$ spherical flow (Flammang 1982, 1984). The self-similar
functions of radius for various physical quantities are exactly those of the
non-relativistic adiabatic spherical flow, and the isodensity surface
of the ADAF is spheroidal rather than disk-like (NY3).
In fact, optically thin two-temperature spherical accretion flow with a
magnetic field produces a spectrum that roughly agrees with that of
ADAF (Melia 1992, 1994).

\subsubsection{Unified description}

Recent works on various types of axisymmetric accretion flow with a
wide range of mass accretion rate make it possible to describe these
seemingly distinct families of accretion flow in a unified way
(Abramowicz et al. 1995; Chen et al. 1995). Each family of solution is
constructed on appropriate
assumptions, and specific name is attached to it, e.g., thin disk, slim
disk, and ADAF. The traditional way of presenting the disk solutions in
a plane of the mass accretion rate, $\dot m$, versus the surface mass
density, $\Sigma$, at a given radius is very revealing, contrast to the
spherical accretion where the global properties like the mass accretion
rate and the total luminosity are used. Different families of solutions
appear as different curves in this plane (Fig. 4).

Since the solutions sometimes depends on the mass of the hole $M$ and the value
of the viscous parameter $\alpha$, we take a specific case of $M = 10
M_\odot$ and $\alpha = 0.1$ (Chen et al. 1995). Figure 4 shows two
disconnected curves, each representing the low-$\Sigma$ flow ({\sl
dotted curve} on the left) and the high-$\Sigma$ flow ({\sl solid
curve} on the right). The former is optically thin to absorption and
the latter optically thick. This shows that there can be more than one
type of accretion flow in a certain range of mass accretion rate. Also,
the former type of accretion flow exists only for $\dot m \ll 1$, while
the latter type is possible for all range of $\dot m$.

The lower (GTD) and middle (RTD) branches of the {\bf S} shaped curve
on the right side of Figure 4 correspond respectively to the classic
gas pressure and radiation pressure dominated thin disk solutions
(Shakura \& Sunyaev 1973). Radiative cooling through the surface of the
disk is dominant over the advective transport. The positive slope of the
gas pressure dominated disk shows the flow being stable, and vice versa
for the radiation dominated disk (Lightman \& Eardley 1974; Shakura \&
Sunyaev 1976). The uppermost branch (SD) of the curve is the slim disk
solution in which the advection of radiation plus gas stabilizes the
radiation pressure dominated flow. The vertical scale height of the
slim disk can be comparable to the radius, especially when $l = e \dot{m}
\rightarrow 1$. Even at this high $\dot m$, the temperature of the flow is
less than $\sim 10^8 \K$, and the spectrum of the emitted radiation is
the superposition of modified blackbodies with different temperatures.
Both the gas pressure dominated and the radiation pressure dominated
thin disks are efficient radiators $e \sim 0.1$ (TD in Figs. 1 \& 3)
while the slim disk can have smaller efficiency $e \lesssim 0.1$ (SD in
Figs. 1 \& 3).

The right branch (SLE in Fig. 4) of the optically thin accretion disk
is the two-temperature hot disk solution of Shapiro et al. (1976). The
slope indicates that the disk is stable to viscous perturbations.
However, it is thermally unstable on much shorter time scales (Pringle
1976; Piran 1978). The flow has different ion and electron
temperatures, and the viscous dissipation energy is readily radiated
through the optically thin flow. Therefore, the radiation efficiency is
always $e \sim 0.1$. Recent work on the stability of SLE solutions
suggests that the advective cooling may affect the thermal
instability (Wu 1997).

The left branch (ADAF in Fig. 4) represents the ADAF of NY1 and Abramowicz et
al. (1995), in which the flow is optically
thin and viscous dissipation energy is mostly advected with the flow,
therefore, stabilizing the flow. The electron temperature of the
flow can reach up to $\sim 10^9 \K$, and the radiation spectrum is that
of the Comptonized bremsstrahlung and synchrotron. The radiation
efficiency of this branch (NY3) is roughly $e \propto \dot m$ (see NY in Fig.
3) and the luminosity $l \propto \dot m^2$ (see NY in Fig. 1) as in the
optically thin spherical accretion (NY3).

There also can be other interesting families of axisymmetric accretion
if processes like pair production (Kusunose \& Mineshige 1992) or
various shocks (Chakrabarti 1996a) exist, which have properties analogous to
the equivalent solutions in the spherical flow. However, preheating, which is
important in the spherical flow, has never
been incorporated in any of the disk solutions shown in Figure 4.

\subsection{Two-dimensional nature of axisymmetric accretion flow}

One important property of the slim or thick axisymmetric accretion flows,
including ADAFs, is their two-dimensional nature. This has been generally
ignored under the assumption that height-integrated equations are good
enough (NY2). However, Park \& Ostriker (1999,
hereafter PO) emphasize that the radiation emitted must interact
with the accreting matter due to the geometrical shape of ADAF . Unlike
the thin disk where the radiation easily escapes the flow through the
vertical surface without any interactions, photons produced at smaller
radii of the flow must escape through the tenuous outer part of the flow
in much the same way as in spherical accretion. By studying the
interaction of radiation with matter in two-dimensional ADAF, they find
that the dynamics and thermal properties of the flow should be
substantially changed in some parts of the parameter space (see also Esin
1997 for one dimensional treatment). Especially, radiative 
cooling and Compton preheating will change the polar region of the
flow dramatically, and under the right conditions winds along the polar
axis will be produced. But PO reached these conclusions based on simple
comparison of various time scales, leaving some interesting questions
unanswered. For example, PO showed that when Compton heating is
negligible, the electrons near the polar axis cannot be maintained at
high temperatures, suggesting the possibility of the cooling of ions.
But answers to what temperature will electrons finally settle to and
whether ions too will cool due to increased coupling can be obtained
only after the energy equations for electrons and ions are properly
solved with all heating, cooling, and energy exchange effects. Also, the
existence of proposed solutions maintained by Compton preheating for
the mass accretion rate at which high temperature solution is not
possible can only be checked by solving the energy equations.

So, in this work we explicitly integrate the ion and electron energy
equations of ADAF with special considerations given to all relevant
radiative cooling, the Compton preheating, and radiative transfer of
Comptonizing radiation. As discussed above, we need to make fundamental
assumptions or simplifications to deal with two dimensional nature of
ADAF. Since we are interested in the two-dimensional thermal properties
of ADAF, we cannot use the usual height-integrated formalism. So we
here adopt the fundamental assumption that the density and the velocity
of the ADAF are described by the self-similar solutions of NY2, regardless
of ion and electron temperature profiles. Since the dynamics of the
flow is controlled by the pressure, and pressure, in turn, is
determined by energetics of electrons and ions, this approach is not
completely self-consistent. In both the work that follows and in NY2 the
pressure is dominated by ions, and since both the cooling and preheating that
we 
treat affect primarily the electrons, the total pressure, which is the quantity
most relevant to the dynamics, differs relatively little between our work and
NY2. In any case, we believe this approach will nicely
complement the extensive work on energetics of ADAF in one-dimensional
formalism (NY3; Narayan, Kato, \& Honma 1997; Chen, Abramowicz,
\& Lasota 1997; Nakamura et al. 1997; Gammie \& Popham 1998; Popham \&
Gammie 1998; Chakrabarti 1996a).

\section{Equations}

\subsection{Self-Similar ADAF}

Viscous accretion flow can be heated by $PdV$ work,
viscous dissipation, and the interaction with radiation.
It cools mostly by emission of radiation.
However, locally, since the flow is moving, advection acts
as loss of gas energy at a fixed point.
The self-similar advection-dominated accretion flow solutions (NY1, NY2)
refer to a family of solutions where viscous heating
is mainly balanced by the advective cooling
with negligible radiative cooling.
The dynamical property of the self-similar ADAF,
e.g., density and velocity, is determined by the viscosity
parameter $\alpha$ and the parameter
$\epsilon^\prime \equiv f^{-1}(5/3-\gamma)/(\gamma-1)$
determined by the adiabatic index $\gamma$ of the gas
and the ratio $f$ of the advective cooling versus the viscous heating. When the
flow is advection dominated, $f\cong1$. 

\subsection{Dynamics}

The big obstacle in investigating
the axisymmetric accretion flow is the two-dimensional nature
of the problem and the global coupling between the radiation and
the gas. The physical state of the gas at one position is
linked with all the other parts of the flow through the transfer
of radiation. Also, the dynamics of the flow has to be determined
self-consistently. It is very difficult at present to solve such
two-dimensional problems
in a truly self-consistent way as was done in spherical accretion
(Shapiro 1973a,b; Park 1990a,b; Nobili et al. 1991).
So we here adopt a fundamental simplifying assumption: the dynamics of the
flow,
i.e., the velocity and density of the gas, will be that of self-similar
solutions (NY2) regardless of the thermal structure of the flow. One can
alternatively consider this an exercise in determining the self-consistency of
the NY2 solution. Thus we take radial velocity to be\footnote{We note,
however, that recent time-dependent 
numerical simulations of two dimensional axisymmetric accretion flows very
near the black hole show diverse dynamical behavior (Igumenshchev \& Abramowicz
1999; Stone et al. 1999).}
\begin{equation}\label{eq:vr}
   v_r(r,\vartheta) = r \Omega_k(r) v(\vartheta),
\end{equation}
the poloidal velocity
\begin{equation}\label{eq:vtheta}
   v_\vartheta = 0,
\end{equation}
the rotational velocity
\begin{equation}\label{eq:vphi}
   v_\phi = r \Omega_k(r) \Omega(\vartheta) \sin\vartheta ,
\end{equation}
where $r\Omega_k(r) = (GM/r)^{-1/2}$.
The density is
\begin{equation}\label{eq:nr}
   n(r,\vartheta)/n_0 = \dot m (r/r_s)^{-3/2} n(\vartheta) ,
\end{equation}
where $n_0 r_s \sigma_T = 1/[2(1+Y_{He})] $ and
$\sigma_T$ Thomson cross section. The free-fall spherical
flow corresponds to $v(\vartheta) = \sqrt{2}$, $\Omega(\vartheta)=0$,
and $n(\vartheta) = 1$.
{\it As noted above, these assumptions would hold only if the pressure field
of the final solution is similar to that of the self-similar one.
We will discuss how self-consistent the final solutions are.}

\subsection{Energy equations}

To get the relevant energy equation in axisymmetric accretion flow,
we extend the relativistic energy equations for the spherical flow
under the radiative heating and cooling
to axisymmetric one (Park 1990a,b).
Although most ADAF calculations are done in a non-relativistic
framework, the difference in temperature between the relativistic
and non-relativistic treatment is so large that we use here the correct
relativistic energy equations. For some parameters, the output luminosity
can be altered by more than a factor of 10.

The ion and electron energy equations are
\begin{mathletters}
\begin{equation}\label{eq:epsilon_i}
   \frac{d\epsilon_i}{dr} - \frac{\varepsilon_i + P_i}{n_i}
   \frac{dn_i}{dr} = \frac{-\Gamma_{vis} + Q_{ie}}{v_r} ,
\end{equation}
\begin{equation}\label{eq:epsilon_e}
   \frac{d\epsilon_e}{dr} - \frac{\varepsilon_e + P_e}{n_i}
   \frac{dn_e}{dr} = \frac{\Lambda_e-\Gamma_e - Q_{ie}}{v_r} ,
\end{equation}
\end{mathletters}
where $\Gamma_{vis}$ is the viscous heating rate for ions,
$\Lambda_e$ and $\Gamma_e$ the cooling and heating rates for
electrons, $\varepsilon_i$ and $\varepsilon_e$ the internal
energies, all per unit volume, $P_i$ and $P_e$ the pressures,
$n_i$ and $n_e$ the number densities, $T_i$ and $T_e$
the temperatures, of ions and electrons, respectively. This form of energy
equations inherently accounts for the advective cooling of both ions and
electrons (see Nakamura et al. [1997] for discussion on advection of
electron energy). 
The energy transfer through Coulomb coupling is described
by the rate $Q_{ie}$ where
\begin{equation}
   Q_{ie} = \frac{3}{2} \frac{m_e}{m_i}
            \sum_i Z_i^2 n_e n_i \sigma_T c (kT_i-kT_e)
            \frac{1+\sqrt{\pi/2}(\theta_e+\theta_i)^{1/2}}
                 {\sqrt{\pi/2}(\theta_e+\theta_i)^{3/2}}
            \ln\Lambda,
\end{equation}
$Z_i$ is ion charge, $m_i$ and $m_e$ ion mass and electron mass, respectively,
$\theta_i \equiv kT_i/m_i c^2$, $\theta_e \equiv kT_e/m_e c^2$,
$\sigma_T$ Thomson cross section, and 
$\ln\Lambda=20$ the Coulomb logarithm.
There exists a better expression
for $Q_{ie}$ in transrelativistic temperatures
(Stepney 1983; Stepney \& Guilbert 1983), but it
is quite costly to evaluate.
The result is not sensitive to the exact functional form of $Q_{ie}$
in the transition regime.

Since electrons usually reach relativistic temperatures,
$\varepsilon_e = (3/2)x_e^\ast n_e k T_e$, where
$x_e^\ast = (2/3)[5\eta\theta_e^{-1} -
(\eta^2-1)\theta_e^{-1}-1]$,
$\eta = K_3(\theta_e^{-1})/K_2(\theta_e^{-1})$ and $K_3$, $K_2$
are the modified Bessel functions (Shapiro 1973a; Park 1990b). The factor
$x_e^\ast$ incorporates the change of the equation of state at
relativistic temperatures and
has an asymptotic value of 1 when $kT_e \ll m_ec^2$ and
2 when $kT_e \gg m_ec^2$. Also, ion energy density
$\varepsilon_i = (3/2)x_i^\ast n_i k T_i$
where $x_i^\ast$ is similarly defined.
The change of the equation
of state at high temperatures can produce a significant difference
compared to a nonrelativistic treatment. Even neglecting
the derivative of $x_e^\ast$ can make an error of factor 2 (Park 1990a).
Direct calculation of $\eta$ is numerically demanding, so we use
a convenient fitting formula instead (Service 1986).

\subsection{Radiative transfer}

In a freely falling spherical accretion flow, the electron scattering
optical depth from radius $r$ to infinity is simply $ \tau_{es}(r) = \dot m
(r/r_s)^{-1/2}$. In ADAF, the flow is moving more slowly than free-fall and
therefore its density and optical depth is higher for the same $\dot m$.
We define the optical depth along a given polar angle $\vartheta$,
\begin{equation}
   \tau_{es}(r,\vartheta) = \dot m (r/r_s)^{-1/2} n(\vartheta),
\end{equation}
which has the highest value along the equator, $\vartheta=\pi/2$. For
$\alpha=0.1$ and $\epsilon'=0.1$, $1.0$, and $10$ ADAF, $n(\pi/2) \sim 30$,
$\sim 50$, and $\sim 400$, respectively (NY2). However, $n(\vartheta)$ along
the pole is $\sim 30$, $\sim 20$, and $\sim 10$ for
$\epsilon'=0.1$, $1.0$, and $10$. Hence the flow remains optically
thin along the pole for $\dot m \lesssim 0.1$.

It is not the intention of this work to rigorously solve the radiative transfer
for the nonspherical flow. We add the emission from all $\theta$
at a given $r$ to get the averaged
luminosity profile $L_X(r)$ and the radiation
temperature $T_X(r)$ which is defined as $4kT_X$ being equal to the
energy-weighted mean photon energy (Levich \& Syunyaev 1971;Park 1990a),
and use these as an approximation to the real
radiation field. This approximation should be a good one for small
$\epsilon'$ flow where the density profile is almost spherical, or for
low $\dot m$ flow where the flow is optically thin. Moreover, the
preheating radiation field at large radius should be well described
by the streaming radiation, and a spherical radiation field should be good
enough approximation. A more advanced treatment would expand the radiation
field into spherical harmonics with the next (herein neglected) term the
quadrupole component (Loeb \& Laor 1992).

Under this simplification, $L_X(r)$ is just the sum of all
frequency-integrated emission, i.e., radiative cooling rate,
inside radius $r$,
\begin{equation}\label{eq:LX}
   L_X(r) = 2\pi \int_{r_{in}}^{r_{out}} 
                 \int_0^\pi \Lambda_{all}(r,\vartheta)
                 \sin\vartheta d\vartheta r^2 dr .
\end{equation}

A far more difficult task is to calculate the Comptonization temperature
$T_X(r)$ without really solving the frequency-dependent radiative transfer
equation. Since we treat every emission as being Comptonized in situ,
the main change of the radiation temperature in radius is due to the addition
of newly produced photons to already existing photons.
If the spectral energy density of existing photons is
$E_X$ and its radiation temperature is $T_X$, incremental addition of
newly produced and locally Comptonized photons of energy density
$dE_S$ and radiation temperature $T_S$ will change the
radiation temperature of the total to
\begin{equation}
   T_X' = \frac{E_X T_X + dE_S T_S}{E_X + dE_S} = T_X + dT_X ,
\end{equation}
where $dT_X$ is the change in $T_X$ within $dr$.
Solving for $dT_X$ gives (Park 1990a)
\begin{equation}\label{eq:TX}
   \frac{d\ln T_X(r)}{d\ln r} = \frac{T_S(r)-T_X(r)}{T_X(r)}
                             \frac{\Lambda_S(r) r}{F(r)} ,
\end{equation}
where $\Lambda_S(r)$ is the $\vartheta$-average of all emissions
and $F(r) = L(r)/4\pi r^2$.
We used the approximate relation $dE_S/E_X = \Lambda_S dr/F$.
The $\vartheta$-averaged radiation temperature, $T_S(r)$,
is calculated as
\begin{equation}
   T_S(r) = \frac{\int_0^\pi 
                  \Lambda_S (r,\vartheta) T_S(r,\vartheta)
                  \sin\vartheta d\vartheta }
                 {\int_0^\pi \Lambda_S (r,\vartheta)
                   \sin\vartheta d\vartheta} .
\end{equation}
In case where both bremsstrahlung and synchrotron emissions
contribute, $\Lambda_S (r,\vartheta) T_S(r,\vartheta)$ is
replaced by $\Lambda_{Cbr}T_S^{Cbr} + \Lambda_{Csyn}T_S^{Csyn}$.

\subsection{Cooling and heating}

We consider atomic cooling, bremsstrahlung, and
synchrotron radiation and their inverse Comptonization
as the main radiative cooling processes on electrons, and
viscous dissipation and Compton scattering as
the main heating processes on ions and electrons, respectively.
Special care has been taken for absorption and inverse Comptonization of soft
bremsstrahlung and synchrotron photons. A detailed description is given in
the Appendix.

\section{Calculation}

\subsection{Method of Calculation}

When the preheating radiation field is negligible, finding the solution
is straightforward. One just integrates the energy equations
(\ref{eq:epsilon_i},b) with 
zero preheating luminosity. However, when the flow is affected by the
preheating radiation, we have to find solutions that satisfy the energy
equations (\ref{eq:epsilon_i},b) but with a not-yet-determined radiation field.

We use the iteration method to build the solution (Park 1990a,b).
First, initial estimates for $L^0(r)=L_0$ and $T_X^0(r)=T_{X,0}$
are guessed for given $M$, $\dot m$, $\alpha$, $\epsilon^\prime$, and the ratio
of the gas pressure to the total (gas plus magnetic) pressure $\beta$.
The gas energy equation (\ref{eq:epsilon_i},b) is
integrated under this $L^0(r)$ and $T_X^0(r)$ for selected values of
$\vartheta$'s. Now we have $T_i(r,\vartheta)$ and $T_e(r,\vartheta)$
at each $(r,\vartheta)$. Normally $r$ is divided into 100 logarithmic
intervals and $\vartheta$ into 10 or 20 intervals from 0 to $\pi/2$.
The angle-averaged luminosity $L^1(r)$ and radiation temperature $T_X^1(r)$
are calculated from this temperature profile (Eqs. [\ref{eq:LX}] and
[\ref{eq:TX}]). 
In general, $L_0(r_{out}) \neq L^1(r_{out})$ and $T_{X,0}(r_{out}) \neq
T_X^1(r_{out})$, and the solution is not self-consistent.
Only for a specific combination of initial $L_0$ and
$T_{X,0}$, will the model be self-consistent in the sense that
the luminosity and the radiation temperature of the iterated solution has
$L^1(r_{out}) = L^0(r_{out})$ and $T_X^1(r_{out}) = T_X^0(r_{out})$.
We search the two-dimensional plane of $L_0$ and $T_{X,0}$ for
solutions that satisfy this self-consistency condition to within the 1 \%
level. 

\subsection{Boundary Conditions}

There are two natural boundaries in accretion problem, inner and
outer ones. At the outer boundary, $r_{out}$, the state of the gas has to
be specified when the gas energy equations are integrated,
and at the inner boundary $r_{in}$,
the state of the radiation field, $L(r_{in})$ and $T_X(r_{in})$, needs
to be specified for equations (\ref{eq:LX}) and (\ref{eq:TX}).

The velocity and density of gas are fixed by equations
(\ref{eq:vr})-(\ref{eq:nr}) for given $\dot m$ and $M$. So the only meaningful
boundary condition is the temperature of electrons and ions.
Due to the strong atomic cooling around $10^4 \K$,
the gas at large radius in spherical flows (except for very low $\dot m$)
is most likely to be at the usual equilibrium temperature $T_e=T_i = T_{eq}
\simeq 10^4 \K$. However, self-similar 
ADAF solutions have $T_e=T_i\sim T_{vir}$ at all radii, where $(5/2) k T_{vir}
\equiv GMm_p/r$. So we will apply both boundary conditions: $T(r_{out}) =
T_{vir}(r_{out})$ or $T(r_{out}) = T_{eq}(r_{out})$. The outer boundary
radius in most calculation is taken to be $10^5 r_s$, which corresponds to
$T_{vir} \sim 10^7 \K$.

We adopt $r_{in} = 3 r_s$ since the radius of the marginally stable orbit
around Schwarzschild black hole is $3r_s$. The luminosity
at $r_{in}$ would be the sum of gravitationally redshifted
emission of radiation between $r_{in}$ and $r_s$, which
we approximate as
\begin{eqnarray}
   L(r_{in}) 
            & \simeq & \int\sin\vartheta d\vartheta 
                   \int_{r_s}^{r_{in}} \Lambda r^3 d\ln r \nonumber\\
            & \simeq & \case{1}{2}\Lambda (r_{in}) r_{in}^3 \ln(r_{in}/r_s),
\end{eqnarray}
and the radiation temperature as
\begin{equation}
   T_X(r_{in}) = \frac{\Lambda_{Cbr}T_S^{Cbr}
                       +\Lambda_{Csyn}T_S^{Csyn}}
                      {\Lambda_{Cbr}+\Lambda_{Csyn}}\Big|_{r_{in}}.
\end{equation}

\section{Results}

The dynamical structure of the flow, i.e., density and velocity, is
fixed by the parameter $\epsilon^\prime$,
which is expected to be between 0 and 1. Esin (1997) argues that
$\gamma = (8-3\beta)/(6-3\beta)$, and $\gamma = 13/9$ for $\beta=1/2$.
This would make $\epsilon^\prime = 1/2$ for $f=1$. But Quataert \&
Narayan (1999) argue that $\gamma$ can be close to 5/3. Also
in relativistic regime $\gamma$ approaches 4/3, and in many of
our solutions $f$ can be different from 1. So we choose
$\epsilon^\prime=1$ as our representative value.
The flow structure for $\epsilon^\prime=0.1$ and that for
$\epsilon^\prime=1$ are not significantly different, whereas
that for $\epsilon^\prime=10$ is rather extreme (NY2). We also choose
$Y_{He}$ to be 0.1. 

\subsection{Un-preheated flow}

First, we searched for solutions that have negligible preheating. 
This is simply achieved by starting the iteration with very small preheating
luminosity. We first choose $T(r_{out})=T_{vir}$ as the outer boundary
condition, which would favor the high temperature solutions.

For the flow onto $M = 10 M_{\sun}$ or $M = 10^8 M_{\sun}$ black holes,
we find solutions when $\dot{m} \la 10^{-4}$ that have near virial ion
temperature everywhere, which validates the self-similar ADAF. 
However, we also find that ions and electrons cool down
$\sim 10^4\K$ at all radii for $\dot{m} \gtrsim 0.04$
regardless of the mass of the black hole, possibly becoming either a cool thin
disk ({\sl dotted line} TD in Figs. 1 \& 3) or a cool spherical flow ({\sl
triangles} in Figs. 1--3). 
This non-existence of hot solutions agrees rather well with one-dimensional
calculation (NY3) despite the different approach:
at $r=10^5 r_s$, the critical mass accretion rate above which
the ADAF of NY3 does not exist is $\dot{m} \simeq 2\times10^{-2}$ (in our
units) for $f=0.3$ and $m=10$.
For the intermediate mass accretion rate $10^{-4} \la \dot m \la 0.04$, both
ions and electrons near the polar axis have $\sim 10^4\K$ while the gas
near the equatorial plane is maintained near virial temperature.
As the mass accretion rate decreases, the low temperature region
around the polar axis decreases: For the mass accretion rate
$\dot{m}=10^{-3}$, only the conical region of the flow within $6\arcdeg$
of the polar axis is cooled, whereas for $\dot{m}=10^{-2}$ that
within $17\arcdeg$ is cooled ({\sl circles} in Fig. 5).
This confirms the fact that electrons around the polar axis will cool down,
and ions too follow electrons due to increased Coulomb
coupling between them (PO). We could not find any case where ions are
kept at high temperature while electrons are cooled down;
once electrons are cool, the Coulomb coupling becomes too strong
to allow a two-temperature plasma. These effect will be more severe for lower
$\alpha$ flow because they have lower radial infall velocity.

Figure 6 shows the typical temperature profiles of ions and electrons
along different $\vartheta$ for the flow with equipartition magnetic field.
{\sl Solid line} shows ion and electron
temperature profiles for $\vartheta=0$ (the polar axis)
in ADAF with $\dot{m}=10^{-2}$. Up to $\vartheta=3\pi/32$
({\sl long-dashed line}), the flow stays near $10^4\K$
({\sl dotted line} for $\vartheta=\pi/32$ and {\sl short-dashed line} for
$\vartheta=\pi/16$).
Because of the very small viscous and adiabatic heating
near the polar axis, electrons cannot be maintained at a high
temperature, and the stronger coupling between ions and electrons
at lower electron temperature makes ions and electrons have
the same temperatures at all radii for these $\vartheta$'s.
Only for large enough $\vartheta$ ({\sl dot-dashed lines} for
$\vartheta=\pi/8$; upper one for ion and lower one for electron) 
ions and electrons
maintain their high temperatures due to higher viscous heating
rate and shorter flow time. As result we expect
a conical region with $\vartheta \lesssim 3\pi/32$ around the polar
axis will collapse while the flow along the equatorial plane
accretes with significant radial velocity at high temperature.
It is likely that the collapse would lead to an empty funnel along the polar
axis. The temperature change across the surface of the funnel is going to be
sudden, and the energy transfer by conduction may lead to further cooling from
the hot part of the flow. But
the existence and the generic shape of the funnel will remain the same. 
Of course this funnel may be filled by a tenuous outgoing wind, but
consideration of this is beyond the scope of the present paper.
However, the temperature profile of the flow near the equatorial plane agrees
with the result from direct integration height-averaged equations with
dynamics (Nakamura et al. 1997; Manmoto et al. 1997). Also, the flow remains at
high temperatures for all values of mass 
accretion rate when the atomic cooling peak is intentionally removed, so it is
clear that the effect we find (a collapse of the polar gas) is due to atomic
cooling. In sum, it appears that for $\dot m > 10^{-4}$ the ADAF solutions
would collapse at the pole, with the funnel possibly filled with a tenuous hot
outgoing wind. 

We now turn to the cases where the temperature at the outer
boundary is equal to the equilibrium temperature
$\sim 10^4\K$.\footnote{This condition would apply when (quasi) spherical flow
at large radii is matched to the inner ADAF.} Because of the strong atomic
cooling peak near $\sim 10^4\K$, electrons stay in this temperature
until very strong heating takes over.
For the same parameters as in $T(r_{out})=T_{vir}$ cases,
we find that the flow is kept at $T_i = T_e \sim 10^4\K$
for all $\vartheta$ when the mass accretion rate
$\dot{m} \ga 2\times10^{-4}$ ({\sl triangles} in Fig. 5).
The result differs significantly from the prior boundary condition
$T(r_{out}) = T_{vir}$ cases.
This implies that only the very low
mass accretion flow will reach the high temperature when
the outer boundary is at $\sim 10^4\K$
and, therefore, the cool thin disk would not switch to hot a ADAF
automatically for intermediate mass accretion rates
unless there is some forcing process.

The luminosity and the radiation temperature (both at the outer boundary) of
lower flow rate ($\dot{m} \la 
2\times10^{-4}$) hot solutions for this boundary condition 
are shown in Figures 7 \& 8, respectively, as {\sl diamonds}. The
dimensionless luminosity $l$ ({\sl diamonds} in Fig. 7) for given dimensionless
mass accretion rate $\dot m$ is about 100 times larger than 
that for the spherical accretion (Shapiro 1973b) because the
density of ADAF is roughly 20--50 times (for $\alpha=0.1$ and
$\epsilon^\prime=1$), depending on 
$\vartheta$, higher than the spherical flow with the same $\dot m$. 
The radiation temperature $T_X$ ({\sl diamonds} in Fig. 8) is
very low due to the copious soft synchrotron photons emitted (Shapiro
1973b). However, it increases as $\dot m$ approaches $10^{-3.7}$ because now
lower temperature electrons mainly cool by bremsstrahlung.
The sudden decrease of $l$ around $\dot m \simeq 2\times10^{-4}$ in
Figure 7 reflects the transition from hot ADAF to cold solutions as discussed
above ({\sl triangles} in Fig. 5). This transition does not depend on the
strength of the magnetic field since it is mainly determined by the atomic
cooling peak near $T \simeq 10^4\K$. For comparison, the detailed ADAF models
for Sgr A${}^\ast$ ($M = 2.5 \times 10^6 \,M_{\sun}$; Narayan et al. 1998) and
NGC 4258 ($M = 3.6 \times 10^7 \,M_{\sun}$; Lasota et al. 1996) are
shown as $\earth$ and $\boxdot$ in Figures 7 and 8, respectively.

The luminosity and the radiation
temperature of the flow without any magnetic field, thereby cooling only by
Comptonized bremsstrahlung, are shown as {\sl stars} in Figures 7 \& 8. Due to
the absence of soft synchrotron photons, the luminosity is much lower while the
radiation temperature is much higher than the flow with magnetic field. An
approximate effective temperature of the disk surface for the gas pressure
dominated thin disk is shown as {\sl dotted line} TD in Figure 8 for
comparison. 

When conditions of self-consistency (with regard to the thermodynamics of the
emitted radiation) are applied to the solutions with $T(r_{out}) =
T_{eq}(r_{out}) \simeq 10^4 \K$, we see from Figure 8 that both the low and
high temperature branches of the ADAF solutions are only possible for $\dot m
\la 10^{-3.7}$. 

\subsection{Preheated flow}

PO suggested the possibility of preheated ADAF solutions,
self-consistently maintained by the Compton preheating
as in spherical accretion flows. These solutions are found by starting from
{\it ad hoc} initial luminosity and radiation temperature high enough
to affect the flow significantly (Park 1990a,b). Two dimensional plane
of the luminosity and the radiation temperature, ($l$, $T_X$),
is searched for the self-consistent values.

As in un-preheated flow, the outer boundary condition is 
important because the physical state of the flow inside depends
on the entropy of the flow at the outer boundary. We have two
possible choices: $T(r_{out}) = T_{eq}(r_{out})\simeq 10^4\K$
or $T(r_{out}) = T_{vir}(r_{out})$. In most cases, the latter
condition is awkward because $T_{vir}(r_{out})$ is not
the equilibrium temperature at $r_{out}$. The flow immediately
adjusts to the equilibrium temperature $T_{eq} \simeq 10^4\K$
and stays at that temperature until much stronger heating
brings the temperature up. The former choice, which is more
natural in usual spherical accretion, unfortunately, is not consistent
with the self-similar ADAF, where the sum of the ion pressure and the electron
pressure is assumed to have 
the virial value. However, this choice would make joining
the ADAF solutions to outer thin disk solution much more
reasonable. So we adopt this boundary condition.

Figures 7 \& 8 show the preheated solutions found in ($\dot{m}$,$l$) and
($\dot{m}$,$T_X$) plane, which may be called preheated ADAF (=PADAF).
The {\sl circles} represent the solutions with an exactly equipartition
magnetic field. We find solutions 
sustained by preheating in the range $6.5\times10^{-3} \lesssim
\dot{m} \lesssim 0.15$, which includes the range in which no
high temperature flow is possible when preheating is not
considered ($\dot m > 0.04$ or $\dot m > 2\times10^{-4}$). 
The luminosity of the preheated flow increases
generally with the mass accretion rate ({\sl circles} in Fig. 7).
However, for $\dot{m}\lesssim0.015$ the luminosity increases
as the mass accretion rate decreases 
because the electron temperature reaches above $\sim 10^9\K$
for this low $\dot{m}$ flow, and synchrotron emission makes increasing
contribution to the total luminosity. The majority of photons
are now low energy synchrotron photons and the radiation
temperature of these solutions decreases as
$\dot{m}$ decreases and $l$ increases. This can be seen quite clearly in Figure
8. The radiation temperature decrease quite rapidly as $\dot m$ becomes smaller
than $\sim 0.015$. Compared to the (un-preheated) ADAF, the PADAF has much
higher 
radiation temperatures because most soft synchrotron photons are absorbed and
inverse Comptonization is stronger. Solutions of NY3 are also shown ({\sl
dotted line}) in Figure 7 for comparison. 

The temperature profile of the typical PADAF solution is shown in Figure 9.
The parameters of the flow are $\dot{m}=0.01$, $M = 10^8 \,M_{\sun}$,
$l = 4.1\times10^{-6}$, and $T_X = 1.2 \times10^9\K$.
At the outer radii, the flow is kept near $10^4\K$.
When the Compton heating becomes significant, the temperature suddenly
jumps above $10^6\K$ first at the pole ({\sl solid line}).
This arises from the classic phase change due to the atomic
cooling peak. Away from the pole, the density is higher, and
the jump occurs at successively smaller radii. Also due to
the faster infall velocity, the transition is smoother.
Once the temperature is above $\gtrsim 10^8\K$,
the Coulomb coupling gets weaker, and electron temperature
deviates from that of the ions. The flow is mostly heated by
Compton preheating by hot photons produced at smaller radii
with added help of adiabatic compression and viscous heating.
The electron temperature increases with varying slope
at $T_e \gtrsim 10^9\K$ by relativistic effects
and then flattens due to highly efficient
Comptonized relativistic bremsstrahlung and Comptonized
synchrotron radiation at this temperature range.

Now we look at a higher $\dot{m}$, higher $l$
solution (Fig. 10 for $\dot m=0.1$, $M=10\,M_\sun$).
Due to the higher density, ions and electrons are well coupled.
The flow in the polar
axis ({\sl solid line}) is heated first at much larger radii
compared to lower $\dot{m}$ flow, and larger
$\vartheta$ flow is heated successively
({\sl dotted line} for $\vartheta=\pi/8$,
{\sl short-dashed line} for $\vartheta=\pi/4$, {\sl long-dashed line} for
$\vartheta=3\pi/8$, and {\sl dot-dashed line} for $\vartheta=\pi/2$).
Note that within the radius range $10^3 < r/r_s < 10^6$, the flow
temperature of the polar region
is much greater than the virial temperature ({\sl thin solid line})
while that for the equatorial plane ({\sl dot-dashed line})
is close to the virial value.
This will produce the preheating instability or wind preferentially
along the pole as suggested in PO because longer infall timescale along the
pole makes this part of the flow more vulnerable to instability while shorter
infall timescale near the equatorial plane stabilizes the flow. We may call
this type of solutions ADAF with polar wind (WADAF).
So, preheating may be yet another process to produce outflow in ADAF
(Xu \& Chen 1997; Blandford \& Begelman 1999; Das 1999; Turolla
\& Dullemond 2000; see also Menou et al. [1999] for the polar wind in neutron
star accretion). Meanwhile, if the mass accretion rate, and accordingly the 
luminosity also, is much higher, all parts 
of the flow will be heated well above the virial value.
Since in spherical flow overheating outside the sonic point produces long term
modulation on the accretion rate and that inside the sonic point produces
short term flaring (Cowie et al. 1978), it is likely that these higher $l$ flow
may develop either a relaxational variability or a flaring variability.

The range of $\dot{m}$ to have equatorial accretion
and the preheated polar outflow simultaneously is rather limited
around $\dot{m} \simeq 0.15$. This range will be wider for
larger $\epsilon^\prime$ and narrower for smaller
$\epsilon^\prime$, because larger $\epsilon^\prime$ flow has a higher
density contrast between the pole and the equator.

When there is no magnetic field, Comptonized bremsstrahlung is
the only cooling process, and the properties of these flow
show more continuous behavior ({\sl crosses} in Figs. 7 \& 8). 
They also exist in a wider $\dot m$ range, $10^{-3.5} \la \dot m
\la 10^{-1}$. 
The slope of $l$ versus $\dot{m}$ changes around $\dot{m}\simeq10^{-2}$
because the electrons can reach near $10^9\K$, and relativistic radiative
process becomes important.

\section{Summary and Discussion}

We have studied the two-dimensional thermal properties
of the ADAF by integrating the energy equations under the
assumption that the density and velocity of the flow
is described by the self-similar solutions
of Narayan \& Yi (1995a). Special considerations are given
to the effects of preheating by Comptonizing hot photons produced
at smaller radii and radiative transfer of these photons,
as well as to the
atomic cooling, Comptonized relativistic bremsstrahlung,
and Comptonized synchrotron radiation.
We find that Compton preheating is important in general.

1. When preheating is not considered, the high temperature flows do not exist
when the mass accretion rate is higher than $\sim10^{-1.5} L_{E}c^{-2}$,
and even below this mass accretion
rate, a roughly conical region around the hole cannot sustain
high temperature ions and electrons. Funnels should exist in these flows unless
$\dot m \equiv \dot M c^2 / L_{E} < 10^{-4}$. If the flow starts at large
radii with a normal equilibrium temperature $\sim 10^4\K$ as in spherical
flows, the critical mass accretion rate becomes $\dot m \sim 10^{-3.7}$, above
which level no self-consistent solutions exist. 

2. Even above this critical mass accretion rate,
the flow can be self consistently
maintained at high temperature if Compton preheating is considered.
These solutions constitute a new branch of solutions as in the case of
spherical accretion flow; these preheated high temperature ADAFs can exist
above the critical mass accretion rate in addition to the usual
cold thin disk. Therefore, Compton preheating could be the mechanism
to trigger the phase change from the thin disk to PADAF---the preheated ADAF.

3. We also find solutions where the flow near the equatorial plane is
normally accreting while that near the polar axis is overheated
by Compton preheating, possibly becoming polar wind. The flow may be called
WADAF---ADAF with a polar wind.

The formalism we have used in this work is not completely satisfactory
in the sense that the dynamics of the flow is prescribed to be ADAF like. In
real flow, changes in thermal properties will induce those in dynamical
properties. Hence, although solving the two dimensional flow structure with
the correct radiative transfer is a daunting task,
understanding the true nature of the ADAF like solutions seems to demand it. We
feel confident, however, that the characteristic solutions for all accretion
rates of interest ($\dot m \ga 10^{-3}$) will be either PADAF or
WADAF type when the outer boundary approaches the equilibrium ($T\sim10^4\K$)
temperature. 

\acknowledgments

We would like to thankfully acknowledge useful conversations with 
E. Quataert, K. Menou, R. Narayan, I. Yi, R. Blandford, X. Chen, and
B. Paczyn\'nski. 
This work was supported by Korea Research Foundation grant
KRF-1999-015-DI0113 and NSF grant 94-24416.

\appendix

\section{Cooling and heating processes}

\subsection{Atomic cooling and bremsstrahlung}

Cooling due to atomic processes and bremsstrahlung is
expressed as a composite formula (Svensson 1982;
Stepney \& Guilbert 1983; Nobili et al. 1991)
\begin{equation}
   \Lambda_{Cbr+atomic} = \sigma_T c \alpha_f m_e c^2 n_i^2
   \Big[
   \big\{ \lambda_{Cbr}(T_e)
          + 6.0\times10^{-22}\theta_e^{-1/2} \big\}^{-1}
         + \left(\frac{\theta_e}{4.82\times10^{-6}}\right)^{-12}
   \Big]^{-1},
\end{equation}
where $\sigma_T$ is the Thomson cross section,
$\alpha_f$ the fine-structure constant, and $n_i$ the number
density of ions.
In part of the flow with high temperature where most of the
radiation is produced, the bremsstrahlung cooling rate
is not much affected by free-free absorption
because most energies are carried by photons with energy
$\sim kT_e$ in bremsstrahlung emission.

The cooling rate due to the Comptonized bremsstrahlung
$\Lambda_{Cbr}$ can be expressed as some average enhancement factor times
the un-Comptonized bremsstrahlung rate
\begin{equation}
   \lambda_{br} =  \left(\frac{n_e}{n_i}\right)(\sum_i Z_i^2)
                   F_{ei}(\theta_e)
                 + \left(\frac{n_e}{n_i}\right)^2 F_{ee}(\theta_e)
\end{equation}
where
\begin{eqnarray}
   F_{ei} & = & 4 \left(\frac{2}{\pi^3}\right)^{1/2}
                \theta_e^{1/2}(1+1.781\theta_e^{1.34})
          \quad\hbox{for}~~ \theta_e < 1 \\
          & = & \frac{9}{2\pi} \theta_e
                \left[\ln(1.123\theta_e+0.48)+1.5\right]
          \quad\hbox{for}~~ \theta_e > 1 \nonumber\\
   F_{ee} & = & \frac{5}{6\pi^{3/2}}(44-3\pi^2)\theta_e^{3/2}
                (1+1.1\theta_e+\theta_e^2-1.25\theta_e^{5/2})
          \quad\hbox{for}~~ \theta_e < 1 \\
          & = & \frac{9}{\pi} \theta_e
                \left[\ln(1.123\theta_e)+1.2746\right]
          \quad\hbox{for}~~ \theta_e > 1 .\nonumber
\end{eqnarray}
Obtaining a precise estimate of the enhancement factor for arbitrary
optical depth, electron temperature, and flow geometry is
a quite formidable task by itself. However, Dermer, Liang,
\& Canfield (1991) provide a convenient, yet reasonable way to deal with
thermal Comptonization. They define a Comptonized
energy enhancement factor $\eta(\nu)$ as the average change
in photon energy between creation and escape
in a flow with an electron scattering
optical depth $\tau_{es}$ and electron temperature $T_e$,
\begin{equation}\label{eq:eta}
   \eta(\nu) = 1 + \eta_1 - \eta_2 \left(\frac{x}{\theta_e}\right)^{\eta_3}
\end{equation}
where $x \equiv h\nu/m_ec^2$,
$P=1-\exp(-\tau_{es})$, $A=1+4\theta_e+16\theta_e^2$,
$\eta_1 \equiv P(A-1)/(1-PA)$, $\eta_3 \equiv -1 -\ln P/\ln A$,
and $\eta_2 \equiv \eta_1/3^{\eta_3}$.
Although this formula is not applicable to certain ranges of $\nu$ and
$\tau_{es}$, it is good enough for general use.

We apply $\eta(\nu)$ to the bremsstrahlung spectrum
\begin{equation}
   \epsilon_{br} (x)
                 d\left(\frac{x}{\theta_e}\right)
   = \Lambda_{br} \exp \left(-\frac{x}{\theta_e}\right)
                 d\left(\frac{x}{\theta_e}\right)
\end{equation}
to get the Comptonized bremsstrahlung cooling rate.
When absorption is not important, photon with
energy $x$ ($x \lesssim 3\theta_e$) would be upscattered to $\eta x$,
and the cooling rate would be
\begin{eqnarray}
   \Lambda_{Cbr} & = & \int_0^{3\theta_e} \eta x
                       \frac{\epsilon_{br}}{x} dx
                     + \int_{3\theta_e}^\infty \epsilon_{br}(x) dx \\
                 & = & \int_0^\infty \epsilon_{br}(x)dx
                     + \int_0^{3\theta_e}(\eta -1)\epsilon_{br}(x) dx.
\end{eqnarray}
For the second integration, we use simpler form of $\epsilon_{br}$,
\begin{eqnarray}
   \epsilon_{br}^0(x) d\left(\frac{x}{\theta_e}\right)
                 & = & \Lambda_{br}
                   d\left(\frac{x}{\theta_e}\right) \quad x \leq \theta_e \\
                 & = & 0 \quad\quad\quad x > \theta_e
\end{eqnarray}
since very little emission exists above $x>\theta_e$.
The resulting cooling rate is
\begin{equation}
   \Lambda_{Cbr} = \Lambda_{br}
                   \left[1 + \eta_1 - \frac{\eta_2}{\eta_3 +1} \right].
\end{equation}

When $\theta_e$ and $\tau_{es}$ increases such that $PA \sim 1$,
most photons are likely to be upscattered to $\sim kT_e$.
Equation (\ref{eq:eta}) has a formal divergence in this saturated
Comptonization regime. If we assume that photons
with energy above $h\nu_{abs}$ are all upscattered to $kT_e$,
the saturated energy enhancement factor for the whole emission is
\begin{eqnarray}
   \Lambda_{Cbr}^{sat} / \Lambda_{br}
   & = & 1 + \int_{x_{abs}}^{\theta_e} \frac{3\theta_e}{x}
         d \left( \frac{x}{\theta_e} \right) \\
   & = & 1 + 3 \ln \left( \frac{\theta_e}{x_{abs}} \right).
\end{eqnarray}
Hence, the Comptonized bremsstrahlung cooling rate in general is
\begin{equation}
   \Lambda_{Cbr}^{sat} = \Lambda_{br}
                 \min \left[1+\eta_2-\frac{\eta_2}{\eta_3+1},
                 1+ 3\ln\big(\frac{\theta_e}{x_{abs}}\big)\right].
\end{equation}

The energy-weighted mean photon energy for the emission
is expressed as
\begin{eqnarray}
   \frac{kT_S^{Cbr}}{m_ec^2} & = & \Lambda_{Cbr}^{-1}
                                   \int_0^\infty (\eta x)^2
                                   \frac{\epsilon_{br}}{x} dx \\
   & \simeq & \int_0^\infty x\epsilon_{br}(x)dx
       + \int_0^{\theta_e} (\eta^2-1) x\epsilon_{br}^0(x) dx \\
   & = & \theta_e \frac{\Lambda_{br}}{\Lambda_{Cbr}}
         \left[ 1 +\eta_1(\eta_1+2)
         -\frac{2\eta_2(1+\eta_1)}{\eta_3+2}
         +\frac{\eta_2^2}{2\eta_3+1} \right].
\end{eqnarray}
Since this expression has divergence near saturated regime,
we put an upper limit $T_S = T_e$ (Wien spectrum),
\begin{equation}
   T_S^{Cbr} = \case{1}{4} kT_e \min\left[4,
               \frac{1 +\eta_1(\eta_1+2)
                  -\frac{2\eta_2(1+\eta_1)}{\eta_3+2}
                  +\frac{\eta_2^2}{2\eta_3+1}}
               {1+\eta_1-\frac{\eta_2}{\eta_3+1}}\right]
\end{equation}
which has a limit value $T_S^{Cbr} = \slantfrac{1}{4} T_e$ when
Comptonization is negligible.

\subsection{Synchrotron}

Even an equipartition magnetic field does not play an important
role in dynamics of spherical accretion (Shapiro 1974).
But synchrotron emission could be the dominant cooling process nevertheless.
In this work we assume a magnetic field of the order
of equipartition strength:
\begin{equation}
   P_m = \frac{B^2}{24\pi} = \beta (P_g+P_m) = \beta P
\end{equation}
where $P_m$ is the magnetic pressure, $P_g$ the gas pressure,
and $P$ the total pressure.

The angle-averaged synchrotron emission by relativistic
Maxwellian electrons is given by (Pacholczyk 1970)
\begin{equation}\label{eq:ep_syn}
   \epsilon_{syn} = \frac{2\pi}{\sqrt{3}} \frac{e^2}{c}
                    \frac{n_e\nu}{\theta_e^2} I'(x_M) d\nu
\end{equation}
where
$x_M \equiv 2\nu/(3\nu_0\theta_e^2)$, $\nu_0 \equiv eB/(2\pi m_e c)$,
and
\begin{equation}
   I'(x_M) = \frac{4.0505}{x_M^{1/6}}
             \left(1 + \frac{0.40}{x_M^{1/4}}
                   + \frac{0.5316}{x_M^{1/2}}\right)
             \exp(-1.8899x_M^{1/3})
\end{equation}
is a fitting formula (Mahadevan, Narayan, \& Yi 1996).

When absorption is not important,
the cooling rate due to the optically thin synchrotron emission is
obtained by integrating the equation (\ref{eq:ep_syn})
\begin{equation}
   \Lambda_{syn}^0 = 213.6 \frac{e^2}{c} n_e \nu_0^2 \theta_e^2.
\end{equation}
In generic accretion flow, a large fraction of the low energy synchrotron
photons can be absorbed either
by synchrotron self-absorption or by free-free absorption,
and its effect should be incorporated along with that of
Comptonization.

If the flow becomes optically thick in absorption at frequency $\nu_{abs}$,
the synchrotron cooling (before Comptonization) can be expressed as
\begin{equation}
   \Lambda_{syn} = f_{syn} \Lambda_{syn}^0
\end{equation}
where
\begin{equation}
   f_{syn} \equiv \int_{x_{abs}}^\infty xI'(x)dx
           \Big/ \int_0^\infty xI'(x)dx
\end{equation}
and $x_{abs} = h\nu_{abs}/m_e c^2$. By this, we assume all photons
above $\nu_{abs}$ escape the gas flow.
The treatment of cooling when the flow is
optically thick to absorption for lower energy photons and thin for
higher energy ones
can be tricky because the absorbed radiation heats the gas and
the slight difference in the radiation field between two different
radii drives the transfer of the energy
while higher energy photons directly escape. Correct calculation
of this would require solving gas energy equations and radiative
transfer equations with frequency dependence. A simpler way would be
to consider only the photons that escape as the source of cooling
(Ipser \& Price 1982), neglecting the diffusive cooling due to the
absorbed photons, which we adopt here (see NY3 for different approach).

For Comptonization of absorbed synchrotron photons, we assume that
photons near the cut-off frequency $\sim \nu_{abs}$ are most affected
by Comptonization, and the cooling rate due to the Comptonization
of synchrotron photons is simply
\begin{equation}
   \Lambda_{Csyn} = \eta(\nu_{abs})f_{syn} \Lambda_{syn}^0.
\end{equation}
Similarly, the radiation temperature of the Comptonized
synchrotron photons, $T_S^{syn}$, is
\begin{equation}
   4 k T_S^{syn} = h \nu_{abs} \eta(\nu_{abs}).
\end{equation}

\subsection{Absorption}

There are two sources of absorption: free-free and
synchrotron self-absorption. At high temperature where most of the
radiation is produced, synchrotron absorption is more important
in general. In this work, we estimate the frequency at which
either absorption is important under the effect of Comptonization,
and use the larger of the two as $\nu_{abs}$.

The difficulty of considering absorption under Comptonization is
that a photon having a big chance of being absorbed
in the absence of Comptonization can be upscattered
and its mean free path for absorption can increase, resulting in
escape before being absorbed. So the effective frequency where
absorption becomes important becomes lower due to Compton scattering.
This would be far more important to synchrotron losses than to
bremsstrahlung because most of emission in synchrotron radiation
is in the tail part of the spectrum and the precise location of
the absorption frequency can change the emission by some exponential
factor.

Dermer, Liang, \& Canfield (1991) estimate the effective
absorption frequency by equating the characteristic
timescale for absorption with that for the energy increase due to
Compton scattering in case of free-free absorption.
We apply this approach to both free-free absorption and
synchrotron self-absorption.

The absorption frequency $\nu_{abs}^{ff}$ for free-free absorption
is estimated from the relation (Dermer, Liang, \& Canfield 1991)
\begin{equation}
   \tau_{ff}(\nu_{abs}^{ff}) \Upsilon = 1
\end{equation}
where $\tau_{ff} = r a_{ff}$ and $a_{ff}$ is the absorption coefficient
\begin{equation}
   a_{ff}(x) = \sqrt{8\pi}
             \frac{\alpha_f^2 \sigma_T r_e^3}{x^2 \theta_e^{3/2}
             \left[ 1 + (8/\pi)^{1/2}\theta_e^{3/2}\right]}
             n_i^2 \bar{g}
\end{equation}
and
\begin{eqnarray}
   \bar{g} & = & (1+3Y_{He})(1+2\theta_2+2\theta_e^2)
                 \ln \left[
                 \frac{4\eta_E(1+3.42\theta_e)\theta_e}{x} \right] \\
           & & +(1+Y_{He}^2)(\frac{3\sqrt{2}}{5}+2\theta_2)\theta_e
                 \ln \left[
                 \frac{4\eta_E(11.2+10.4\theta_e^2)\theta_e}{x} \right],
                 \nonumber
\end{eqnarray}
$\alpha_f$ the fine structure constant, and $r_e = e^2/m_ec^2$
the electron radius (Svensson 1984). This expression is
valid for $x\ll\theta_e$ and we used
$\sqrt{\pi/2}\theta_e^{1/2}[1 + (8/\pi)^{1/2}\theta_e^{3/2}]$
to approximate $\exp(1/\theta_e)\hbox{K}_2(1/\theta_e)$.
The factor $\Upsilon$ includes the effect of Comptonization,
\begin{eqnarray}
   \Upsilon & = & \frac{1}{\min[1,8\theta_e] \tau_{es}}
            \quad\hbox{if}~~y>1~~\hbox{and}~~\tau_{es}>1 \\
            & = & 1 + \xi \tau_{es} \quad\quad\quad\hbox{otherwise}
\end{eqnarray}
where $\xi = 1/3$ is the geometry factor and
$y \equiv (4\theta_e+16\theta_e^2)\max[\tau_{es},\tau_{es}^2]$
is the usual Compton parameter.

The absorption frequency $\nu_{abs}^{syn}$ for synchrotron
self-absorption is similarly calculated (Svensson 1984;
Dermer, Liang, \& Canfield 1991),
\begin{equation}
   \tau_{syn}(\nu_{abs}^{syn}) \Upsilon = 1
\end{equation}
where
\begin{equation}
   \tau_{syn} = \frac{1}{4\sqrt{3}}
                \frac{e^2 c n_e r}{\nu kT_e \theta_e^2}
                I'(x_M)
\end{equation}
(Ipser \& Price 1982; Mahadevan, Narayan, \& Yi 1996).

Then, the effective absorption frequency $\nu_{abs}$ is set
to be the maximum of $\nu_{abs}^{ff}$ and $\nu_{abs}^{syn}$.

\subsection{Viscous heating}

The ion heating rate per unit volume under $\alpha$-viscosity description
(Shakura \& Sunyaev 1973) is (NY2)
\begin{equation}\label{eq:GCompton}
   \Gamma_{vis} = \alpha\rho c_s^2 \Omega_K
                  \left[ 3v^2(\vartheta) + \case{9}{4}\Omega^2(\vartheta)
                         \sin^2\vartheta
                + \big(\frac{\partial v(\vartheta)}{\partial\vartheta}\big)^2
                + \sin^2\vartheta
                  \big(\frac{\partial \Omega(\vartheta)}{\partial\vartheta}\big)^2
                  \right].
\end{equation}

\subsection{Compton heating}

We now turn to process which can be dispositive in disrupting accretion flow.
Electrons in outer regions, $r/r_s \sim 10^4 \rightarrow 10^5$,
can be heated by high-energy photons produced at inner hot region, 
$r/r_s \sim 10^1 \rightarrow 10^2$, of the flow. 
We use Compton heating rate
from the Kompaneets equation (Levich \& Syunyaev 1971;
Rybicki \& Lightman 1979),
\begin{equation}
  \Gamma_{Compt} = n_e \sigma_T \frac{4k[T_X(r)-T_e(r)]}{m_ec^2}
                   \frac{L_X(r)}{4\pi r^2}.
\end{equation}
The information on the preheating radiation spectrum is incorporated in
the angle-averaged quantities,
$T_X(r)$ and $L_X(r)$, which are calculated from the equations
(\ref{eq:LX}) and (\ref{eq:TX}) along with corresponding auxiliary equations.

\clearpage

\begin{figure}
\plotone{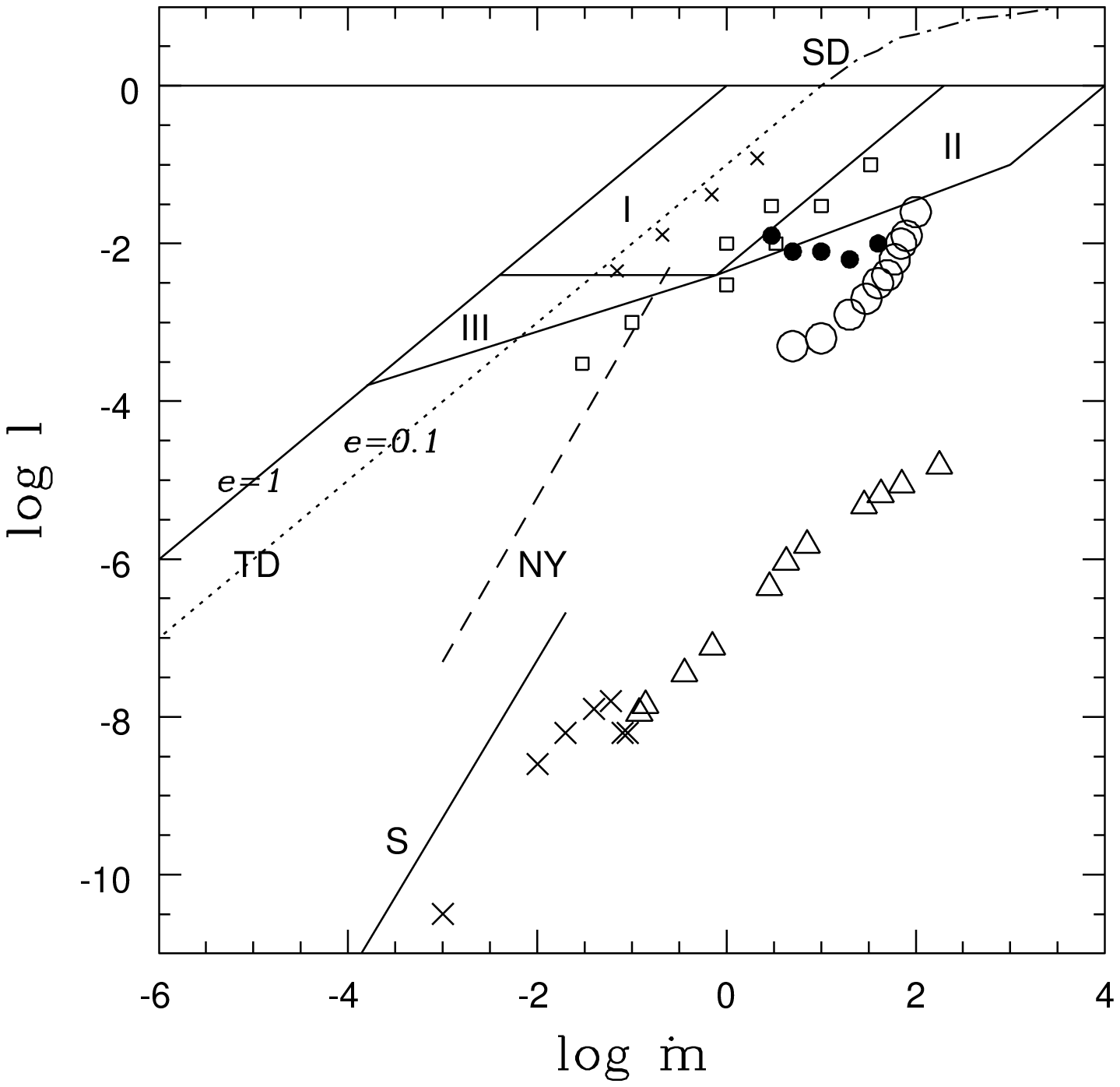}
\caption{Spherical and disk accretion flow solutions in ($l$, $\dot m$)
plane [$l \equiv L/L_{E}$, $\dot m \equiv \dot M c^2 /L_{E}$].
{\sl Solid line} S 
represents low $\dot m$ spherical accretion solutions of Shapiro (1973a), {\sl
large crosses} are those of Park (1990b). Low temperature solutions of high
$\dot m$ spherical 
accretion are {\sl triangles} (Nobili et al. 1991) and high temperature
solutions are {\sl large open circles} (Park 1990b). Accretion disk solutions
appear as {\sl dotted line} TD for thin disk, {\sl dot-dashed line} SD for slim
disk, and {\sl dashed line} NY for ADAF (NY3). Spherical accretion solutions
with electron-positron pairs are {\sl small filled circles} (Park \& Ostriker
1989), those with the magnetic field reconnection {\sl small squares}, and
those with preheating shocks {\sl small crosses} (Chang \& Ostriker
1989).}
\end{figure}

\begin{figure}
\plotone{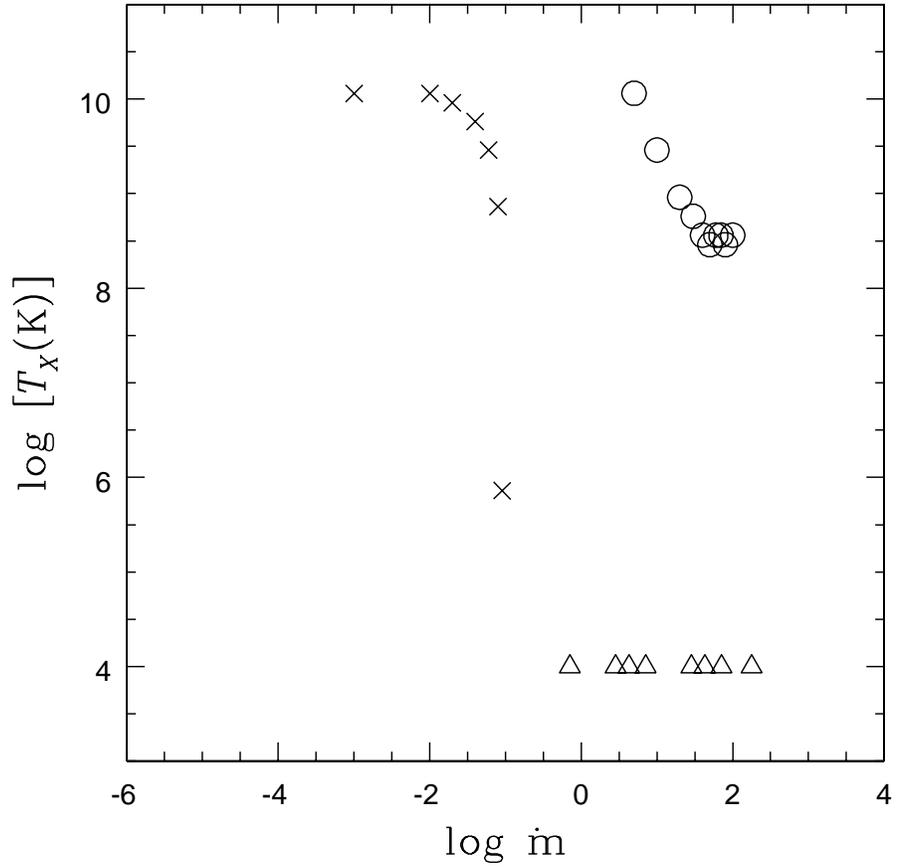}
\caption{The radiation temperature $T_X(r_{out})$ (in K) of the self-consistent
spherical solutions. Symbols are the 
same as in Figure 1. Note that the non magnetic high and low temperature
solutions ({\sl crosses} and {\sl triangles}) are quite distinct on this plane,
but continuous in Figure 1. The separate branch of the self-consistent high
temperature, high-luminosity flows ({\sl circles}) exists also in the ADAF case
(PADAF).} 
\end{figure}

\begin{figure}
\plotone{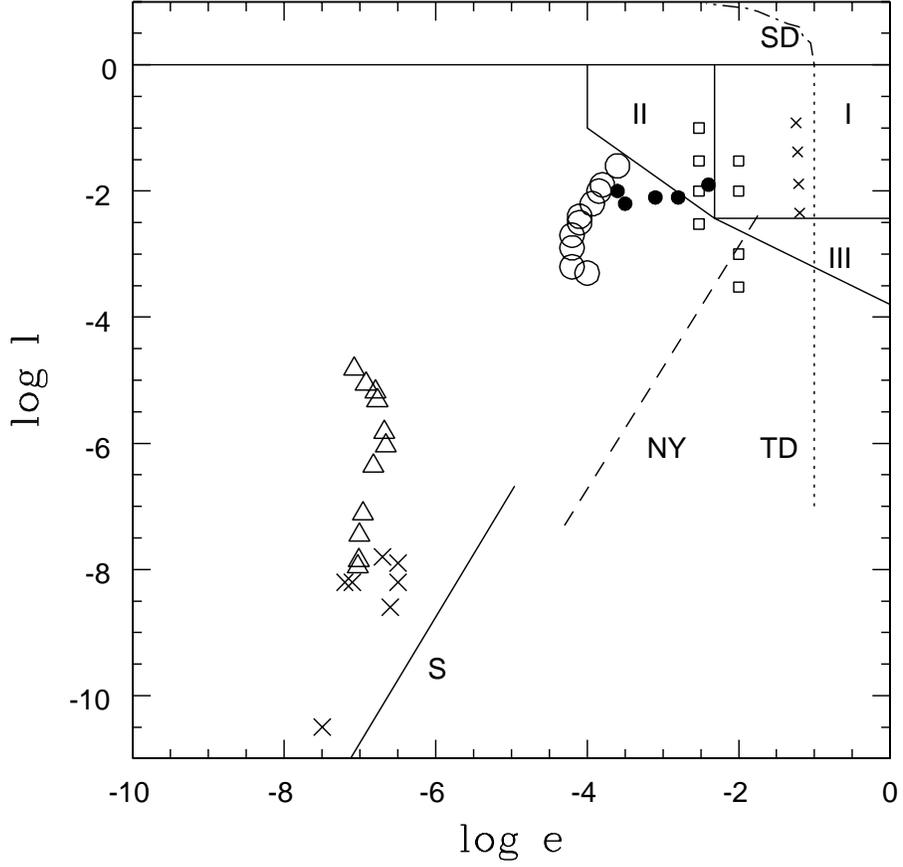}
\caption{Spherical and disk accretion flow solutions in ($l$, $e$)
plane [$e \equiv L/\dot M c^2$]. 
Symbols are the same as in Figure 1. Note that all of the physically 
thick, self-consistent solutions expected or known to be stable ({\sl crosses},
{\sl triangles}, and {\sl circles}) have an efficiency $\la 10^{-4}$. }
\end{figure}

\begin{figure}
\plotone{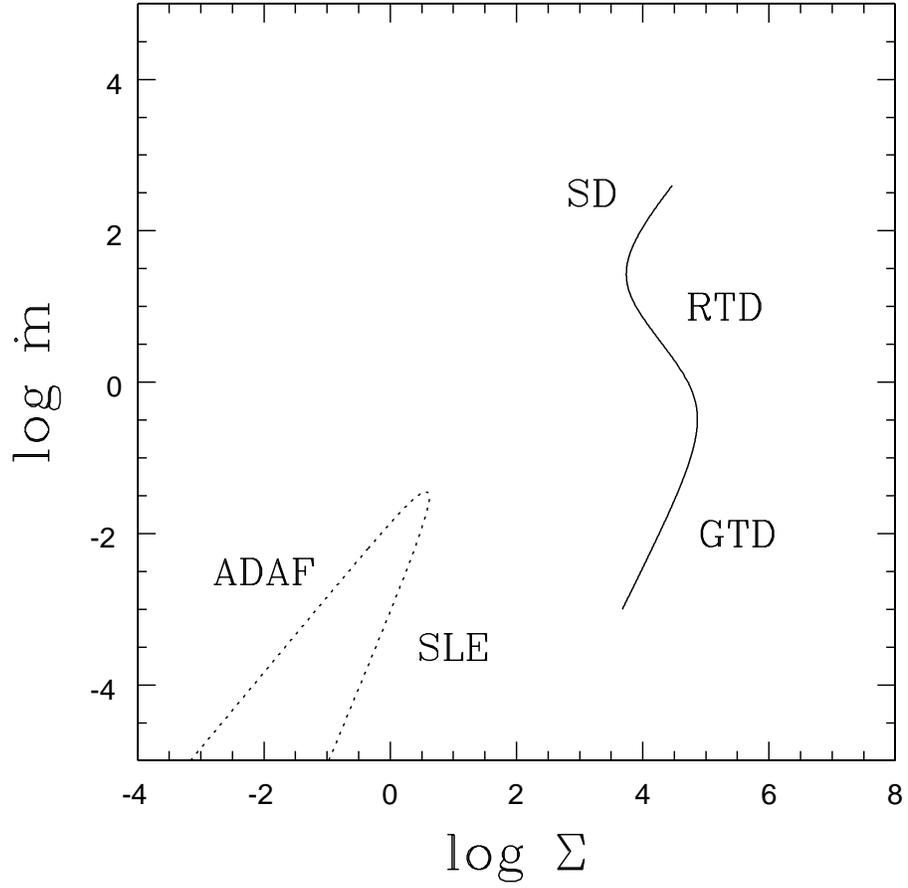}
\caption{Various accretion disk solutions in accretion rate $\dot m$ versus
surface mass density $\Sigma$ (in arbitrary units) plane: {\sl ADAF} for ADAFs,
{\sl SLE} for high temperature, geometrically thin disk solutions, {\sl SD} for
slim disk solutions, {\sl RTD} for radiation pressure dominated thin disk
solutions, and {\sl GTD} for gas pressure dominated thin disk solutions.}
\end{figure}

\begin{figure}
\plotone{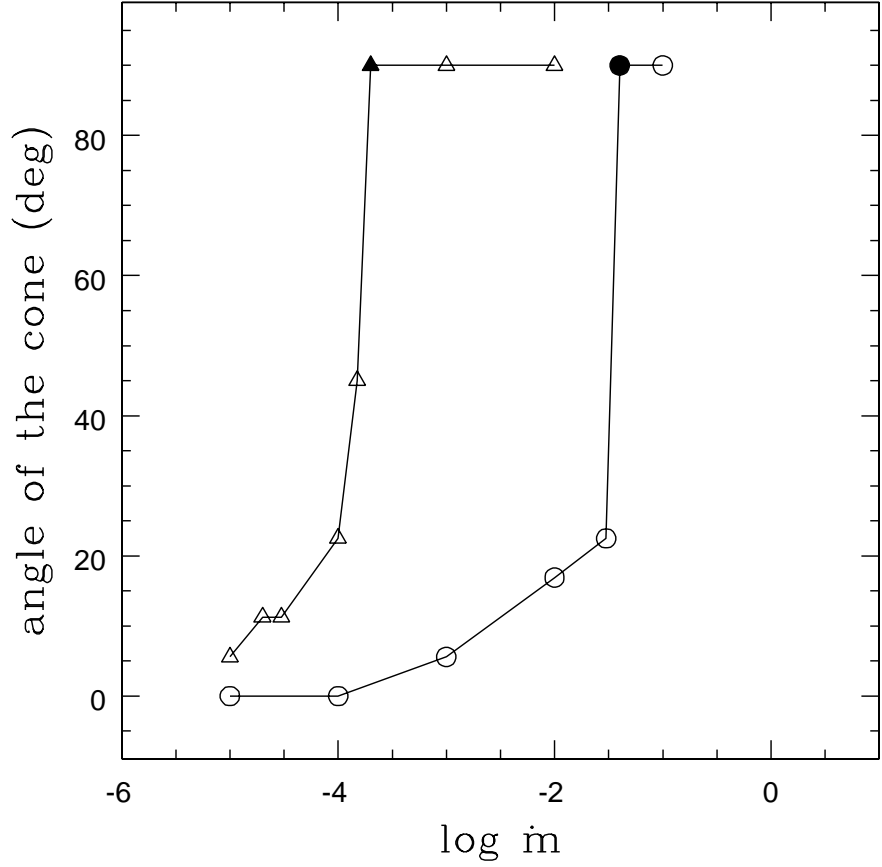}
\caption{For the branch of the ADAF solutions for which preheating is not
important, we plot the angle of the conical region (in degree) around the polar
axis where the temperature of the flow is $\sim 10^4\K$ as a function of $\dot
m$. {\sl Circles} for boundary condition $T(r_{out}) = T_{vir}(r_{out})$ and
{\sl triangles} for $T(r_{out}) = T_{eq}(r_{out})$. With the specified cone
angle pressure balance cannot be maintained and a conical hole would
appear. Thus no solution is possible for $\dot m$ above the limiting points
indicated by {\sl the filled symbols}.}
\end{figure}

\begin{figure}
\plotone{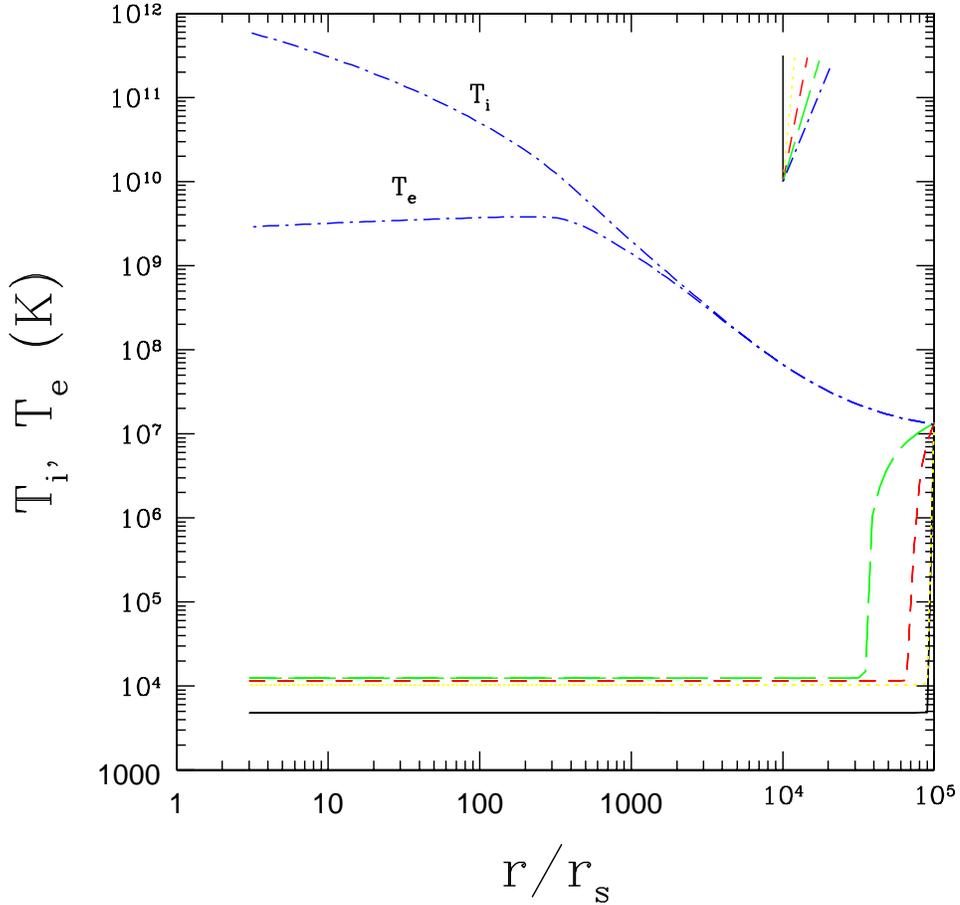}
\caption{Ion ($T_i$) and electron ($T_e$) temperature profiles of $\dot m
=0.01$, $M = 10^8 \,M_{\sun}$ {\sl un-preheated} flow with equipartition
magnetic field for given $\vartheta$'s. 
{\sl Solid line} shows $T_i(r;\vartheta=0) = T_e(r;\vartheta=0) $, 
{\sl dotted line} for $\vartheta=\pi/32$, 
{\sl short-dashed line} for $\vartheta=\pi/16$, 
{\sl long-dashed line} for $\vartheta=3\pi/32$, and
{\sl dot-dashed lines} for $\vartheta=\pi/8$ (upper one for ion and lower one
for electron.) The profiles for larger $\vartheta$'s are almost same as
those of $\vartheta=\pi/16$ case.}
\end{figure}

\begin{figure}
\plotone{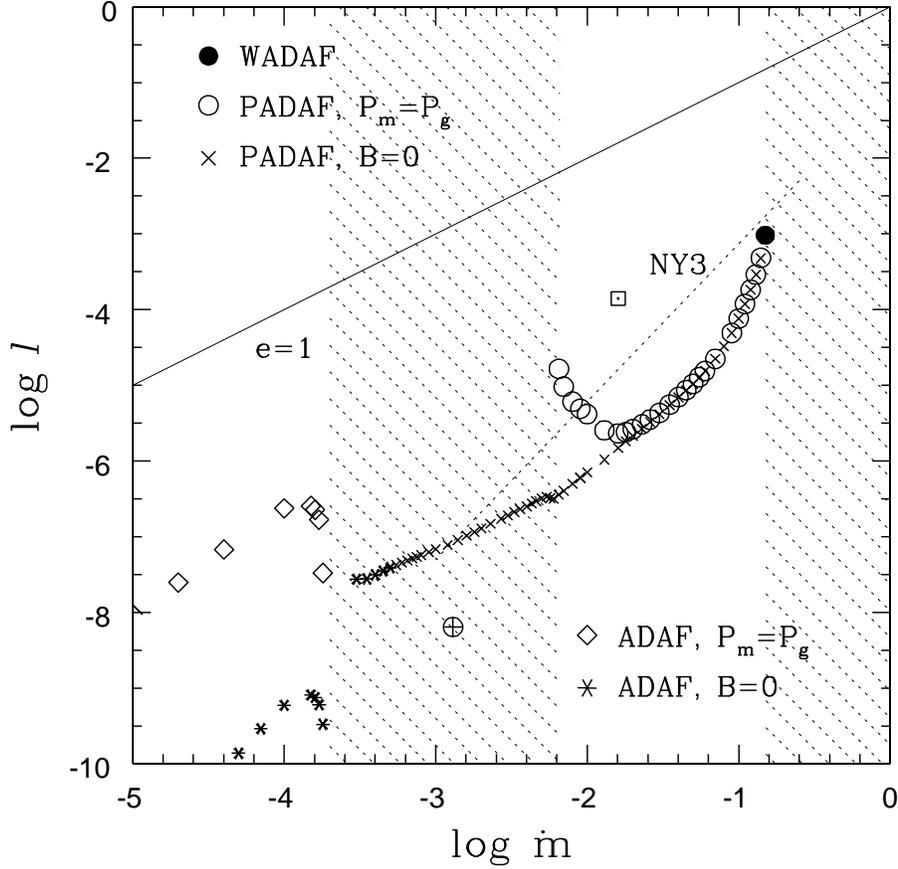}
\caption{The dimensionless luminosity $L(r_{out})/L_{E}$ of solutions found
in this work for accretion onto $10 \,M_{\sun}$ black hole. 
{\sl Circles} and {\sl diamonds} are those with 
equipartition magnetic field, and {\sl crosses} and {\sl stars} are those
without magnetic field. {\sl Diamonds} and {\sl stars} are ADAFs where
preheating is negligible while {\sl circles} and {\sl crosses} are ADAFs
maintained by preheating (PADAF). {\sl Filled circle} represent ADAF with polar
wind (WADAF). One dimensional ADAF solutions (NY3) are
shown as {\sl dotted line}. Detailed ADAF models for Sgr A${}^\ast$ ($M = 2.5
\times 10^6 \,M_{\sun}$) and NGC 4258 
($M = 3.6 \times 10^7 \,M_{\sun}$) are shown as $\earth$ and $\boxdot$, 
respectively. ADAFs with equipartition magnetic field are not permitted 
in the shaded region.}
\end{figure}

\begin{figure}
\plotone{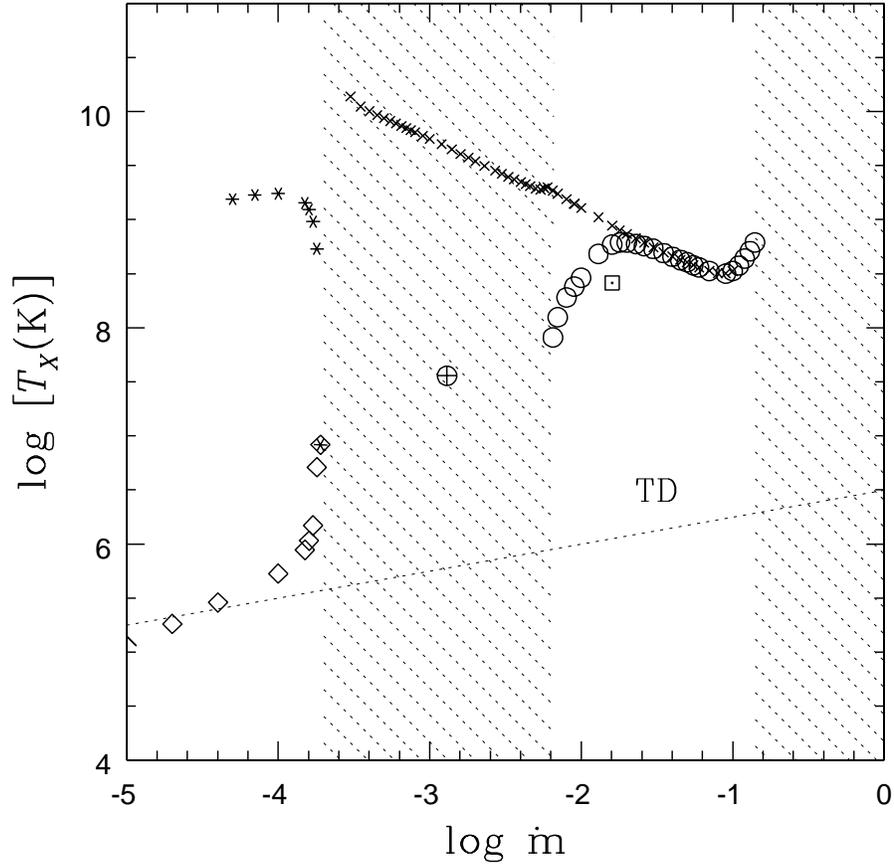}
\caption{The radiation temperature $T_X(r_{out})$ (in $\K$) of solutions found
in this work for accretion onto $10 \,M_{\sun}$ black hole with $T(r_{out})
\simeq 10^4 \K$. Symbols are same as in Figure 7. The effective temperature of
the thin disk surface is shown as {\sl dotted line} TD. Note that without
preheating there are no self-consistent ADAF solutions with $\dot m \ga
10^{-3.7}$. } 
\end{figure}

\begin{figure}
\plotone{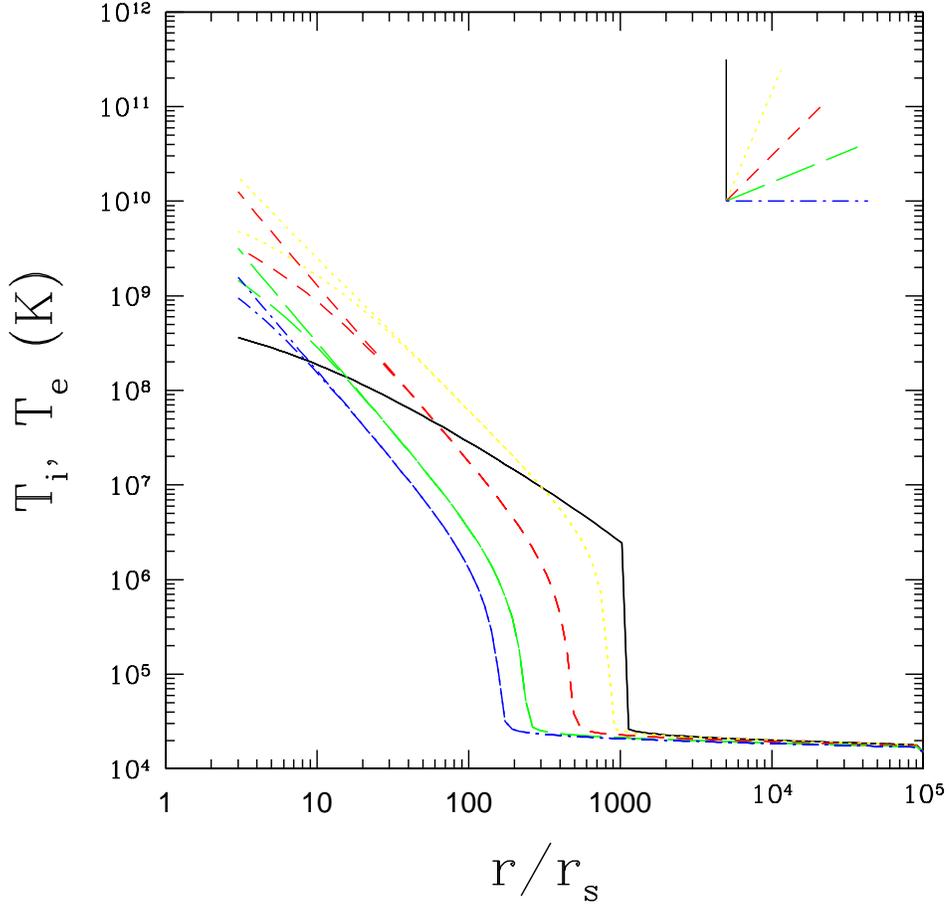}
\caption{Ion ($T_i$) and electron ($T_e$) temperature profiles of PADAF with
equipartition magnetic field for given $\vartheta$'s: 
$\dot m =0.01$ and $M = 10^8 \,M_{\sun}$
($l=7.6 \times 10^{-7}$, $T_X = 1.2 \times 10^9 \K$). 
Now, {\sl solid line} shows $T_i(r;\vartheta=0) = T_e(r;\vartheta=0) $, 
{\sl dotted lines} for $\vartheta=\pi/8$, 
{\sl short-dashed lines} for $\vartheta=\pi/4$, 
{\sl long-dashed lines} for $\vartheta=3\pi/8$, and
{\sl dot-dashed lines} for $\vartheta=\pi/2$. Upper curves for ion temperature
and lower ones for electron temperature.}
\end{figure}

\begin{figure}
\plotone{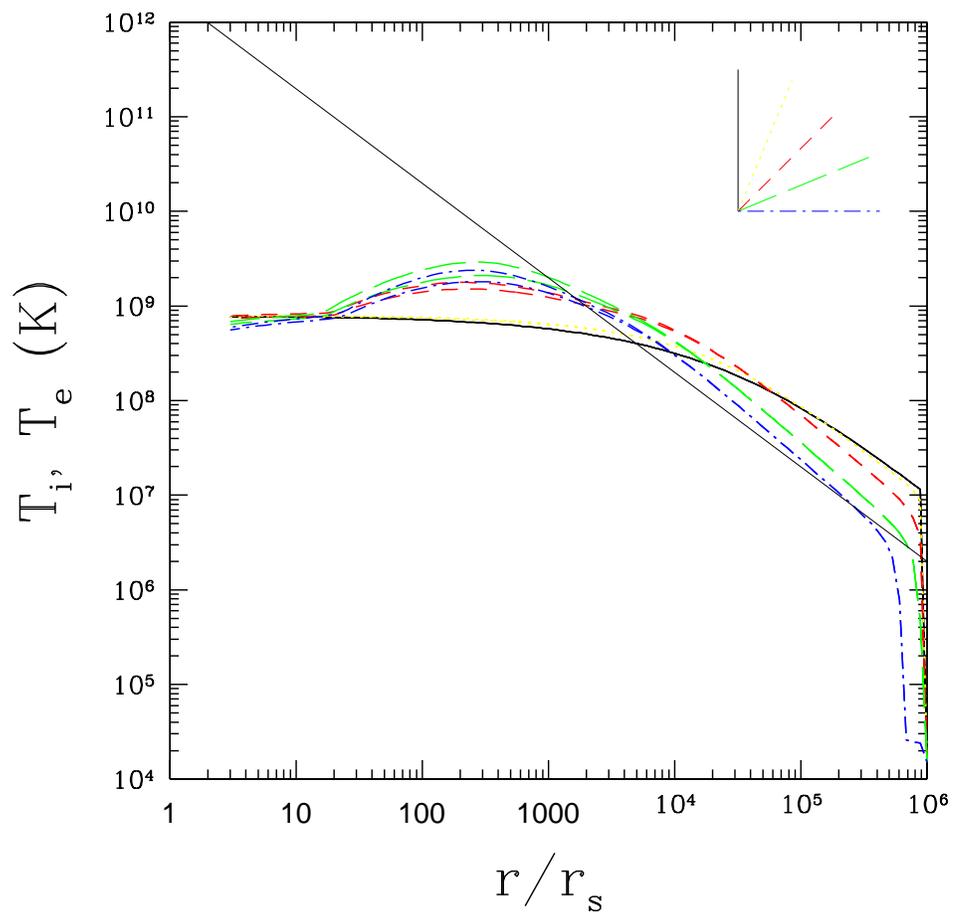}
\caption{Same as Figure 9 but for $\dot m =0.15$, $M = 10 \,M_{\sun}$ ($l=1.0 
\times 10^{-3}$, $T_X = 8.2 \times 10^8 \K$) WADAF. {\sl Thin solid line}
shows the virial temperature.}
\end{figure}

\end{document}